\documentclass[journal,twoside]{IEEEtran}
\newcommand{\FigWidth}{8.8cm}

\usepackage[cmex10]{amsmath} % Only type 1 fonts will utilized at all point sizes, avoiding bitmaps
\usepackage{amssymb}
\usepackage{amsmath}
\usepackage{array}
\usepackage{balance}
\usepackage{booktabs}
\usepackage[normal,footnotesize]{caption}
\usepackage{cite}
\usepackage{color}
\usepackage{epstopdf}
\usepackage{graphicx}
\usepackage{latexsym}
\usepackage{psfrag}
\usepackage{setspace}
\usepackage{siunitx}
\usepackage[caption=false,font=footnotesize]{subfig} % Prevents automatic loading of Axel Sommerfeldt's caption.sty which will override IEEEtran.cls
\usepackage{textcomp}
\usepackage{url}
\newcommand{\insertonefig}[5]
{
\begin{figure}[!ht]
\begin{center}
\includegraphics[angle=#1,width=#2,clip,keepaspectratio]{#3.pdf}
\end{center}
\caption{#5}
{#4}
\end{figure}
}

\newcommand{\inserttwofigsV}[6]
{
\begin{figure}[!ht]
\begin{center}$
\begin{array}{c}
\includegraphics[angle=#1,width=#2,clip,keepaspectratio]{#3.pdf} \\
\text{(a)} \\
\linebreak[4] \\
\includegraphics[angle=#1,width=#2,clip,keepaspectratio]{#4.pdf} \\
\text{(b)} \\
\end{array}$
\end{center}
\caption{#6}
{#5}
\end{figure}
}

\newcommand{\noi} {\noindent}

\newcommand{\cbL}{\mbox{\boldmath ${\cal L}$}}
\newcommand{\cbM}{\mbox{\boldmath ${\cal M}$}}
\newcommand{\cbN}{\mbox{\boldmath ${\cal N}$}}
\newcommand{\cbT}{\mbox{\boldmath ${\cal T}$}}

\begin{document}

% Paper title

\title{PLC-to-DSL Interference:\\Statistical Model and Impact on DSL (Revised)}

% For Drafts and Transactions
\author{{Stefano~Galli,~\IEEEmembership{Fellow,~IEEE,}
        Kenneth J. Kerpez,~\IEEEmembership{Fellow,~IEEE,}\\
        Hubert Mariotte,
        Fabienne Moulin}% <-this % stops a space

\thanks{Stefano Galli (sgalli@ieee.org) is a consultant in New York City, USA.}% <-this % stops a space
\thanks{Ken Kerpez (kkerpez@assia-inc.com) is with ASSIA, Redwood City, California, USA.}% <-this % stops a space
\thanks{Hubert Mariotte (hubert.mariotte@orange.com) and Fabienne Moulin (fabienne.moulin@orange.com) are with Orange Labs in Lannion, France.}% <-this % stops a space

} % end of \author

% The paper headers - \MakeLowercase{...}
\markboth{Submitted to IEEE JSAC Special Issue -- 15 July 2015/Revised 4 December 2015}{Galli \MakeLowercase{\textit{et al.}} ``PLC-to-DSL Interference: Statistical Model and Impact on DSL (Revised)''}

% Make the title area
\maketitle

%\begin{singlespace}
\begin{abstract}
%\boldmath
Newly available standards for broadband access using Digital Subscriber Lines (DSL) have a high degree of spectrum overlap with home networking technologies using broadband Power Line Communications (BB-PLC) and this overlap leads to Electromagnetic Compatibility issues that may cause performance degradation in DSL systems. This paper studies the characteristics of measured PLC-to-DSL interference and presents novel results on its statistical characterization. The frequency-dependent couplings between power line cables and twisted-pairs are estimated from measurements and a statistical model based on a mixture of two truncated Gaussian distributions is set forth. The proposed statistical model allows the accurate evaluation of the impact of BB-PLC interference on various DSL technologies, in terms of both average and worst-case impacts on data rate. This paper further provides an extensive assessment of the impact of PLC-to-DSL interference at various loop lengths and for multiple profiles of Very-high speed Digital Subscriber Lines 2 (VDSL2), Vectored VDSL2 (V-VDSL2), and G.fast. The results of this paper confirm that the impact of PLC interference varies with loop length and whether Vectoring is used or not. Furthermore, the average impact is found to be generally small but strong couplings can lead to substantial degradation of DSL.
\end{abstract}
%\end{singlespace}

% Note that keywords are not normally used for peer review papers.

\thispagestyle{empty}
%\newpage

\section{Introduction}
\IEEEPARstart{O}{ver} the past two decades, broadband access technologies have seen an increasing level of sophistication across a variety of media, allowing consumers to enjoy an unprecedented growth in access speed \cite{Book:BBaccess-Galli2014}. The most widespread broadband technology in the world is Digital Subscriber Lines (DSLs), a family of technologies that provides broadband access over the local telephone network \cite{Book:StaCioSil1999}. Roughly two-thirds of broadband subscribers worldwide are DSL subscribers, and there are more new DSL subscribers each month than new subscribers for all other access technologies combined \cite{Link:BBF-PR-2013}.

One of the recent DSL technologies being deployed today by operators is Very-high speed Digital Subscriber Lines 2 (VDSL2) \cite{STD:ITU:VDSL2-2015}. VDSL2 operates on frequencies from 138 kHz up to 17 MHz or, more recently, 35 MHz \cite{STD:ITU:VDSL2-2015a1}, and transmits at speeds up to a few hundred Mbps and over distances up to 1.5 km. VDSL2 can still be limited by far-end crosstalk (FEXT) which can severely degrade VDSL2 data rates in dense deployments. Vectoring is a technique that allows canceling self-FEXT in both directions under certain deployment conditions. Vectoring has been standardized for use with VDSL2 \cite{STD:ITU:Vectoring-2010, OksmScheClau2010}.

The latest DSL technology that has been standardized in the ITU-T is G.fast (Fast Access to Subscriber Terminals) \cite{STD:ITU:Gfast-2014}. G.fast aims at providing ultra-high speeds over copper twisted pairs up to and sometimes even exceeding speeds of 1 Gbps. Initial requirements foresaw G.fast operating up to 250 meters, but more recently some operators are interested in longer distances such as 400-500 meters. G.fast uses Vectoring to cancel self-FEXT and is specified to operate over frequencies up to 106 MHz \cite{TimmGuenNuzm2013Gfast}. A 212-MHz profile to be used with Vectoring implemented with a non-linear precoder is in initial stages of standardization in the ITU-T.

As the speed of access technologies keeps growing, home networking technologies will have to keep up in terms of data rate. A good candidate for high speed home networking is Broadband Power Line Communication (BB-PLC) as it leverages the existing power cables in the home and has today reached a good technological maturity \cite{BigGalLee2006, Book:PLC-FerreiraLampe-2010, Book:PLC-MIMOandNB-2014, GalliScagWang2011ProcIEEE}. Up to a few years ago, PLC data rates of a few hundred Mbps were available using technologies operating in the 1.8-30 MHz band and based on the IEEE 1901 Broadband Powerline Standard \cite{GalLog2008} and the ITU-T G.hn standard \cite{OksGal2009}. Recently, new PLC technologies using frequencies up to 86 MHz with MIMO capabilities have also been specified, e.g. HomePlug AV2 \cite{YongeAbadHPAV2} and the ITU-T MIMO G.hn \cite{STD:ITU:GhnMIMO}. These advanced PLC technologies achieve data rates between 1-2 Gbps.

In PLC, signals are transmitted over existing power cables which are neither shielded nor balanced, especially in the home. As a consequence, unwanted radiated emissions occur when PLC is used and this phenomenon can cause Electro-Magnetic Compatibility (EMC) issues with other services that operate in the same frequency band of PLC. The issue of coexistence between technologies sharing the same spectrum is well known in wireless communications, e.g. see \cite{GolmieWiFi2001interf, GalFamKod04}. However, while PLC EMC issues have been extensively studied for the case of PLC interference to radio services, little is known today about the effects of PLC interference on other wired communications services - except for the case of self-PLC interference between non-interoperable PLC devices sharing the same band \cite{GalKurOhu2009}.

Since over the past decade the amount of frequency overlap between BB-PLC and DSL has grown considerably, an important example of such EMC issues is the interference that BB-PLC may create to DSL technologies such as VDSL2, Vectored VDSL2 (V-VDSL2), and G.fast. Furthermore, the new Vectoring capabilities of the most recent DSL technologies makes DSL more susceptible to PLC interference (as well as to any other in-home noise) because the time-invariant ``blanket'' of crosstalk that once covered most in-home noises is now gone thanks to Vectoring.

The issue of whether PLC can create harmful interference to DSL or not has been addressed for example in \cite{TR:OfCom:CompatVDSL-PLT, TR:ETSI:PLC-VDSLcoex, MoulinPerZed2010, PrahoTliMou2011, MaesTimm2011}, see also see the excellent bibliographic review in \cite{AliMessDSLandPLCcoex}. The goal of available studies was mostly to understand the general effect of PLC interference on DSL, especially its dependency on cable topologies (e.g., length, separation, load, twisting) and the effectiveness of interference mitigation techniques. However, to the best of the Authors' knowledge, statistical models for evaluating PLC-to-DSL interference are not publicly available and the impact of PLC interference on DSL has not yet been assessed accurately.

This paper reports results from a vast measurement campaign of PLC-to-DSL interference. Orange performed the measurement campaign, reporting a first set of results in \cite{MoulinPerZed2010, PrahoTliMou2011}, with the focus being the impact on PLC interference of topological aspects (length, separation, etc.). This paper uses that data to develop a novel statistical model for PLC-to-DSL interference couplings in the frequency domain\footnote{~The term ``coupling'' is used in this paper in the same way it is normally used in the DSL literature for crosstalk \cite{GalKer02}, i.e. the attenuation in dB at a certain frequency between the excitation signal on a first wire (in our case, the PLC signal on the power line) and the interference observed on a second wire (in our case, the twisted-pair).}. Furthermore, an analysis of the impact of PLC-to-DSL interference on three wideband DSL technologies (VDSL2, V-VDSL2, G.fast) is also presented based on couplings either estimated from measurements or generated according to the model proposed here. The impact on data rate was assessed via computer simulation, using also measured loop responses and measured FEXT.

Achieving a good understanding of the impact that BB-PLC interference may have on DSL is important today as proposals for relaxing the PLC PSD above 30 MHz have been recently discussed in CENELEC and a decision is expected by the end of 2015. Furthermore, following the concern expressed by several telecom operators, the ITU-T and ETSI have started to work on specifying techniques that allow mitigation of PLC-to-DSL interference -- see for example the ongoing ITU-T Q4/Q18 joint effort on Recommendation G.9977 (Mitigation of interference between DSL and PLC).

Initial results were first presented in March 2014 at the ITU-T DSL Standardization Working Group \cite{ITU:PLC-DSL-Interference:teleconf} and then at the IEEE ISPLC 2015 conference \cite{KerpezGalliMariISPLC2015, GalliKerpezMariISPLC2015}. In this paper additional new results are presented:

\begin{enumerate}
  \item An additional 113 measured traces were used to refine previous results.
  \item More accurate statistical models are presented here as they are tailored to specific DSL bandplans.
  \item A new VDSL2 profile currently under standardization that specifies a 35 MHz profile is also considered.
  \item A new G.fast profile operating over a 212 MHz bandwidth with a non-linear GDFE-based pre-coder is also considered.
  \item A new and simple model for the 50\% and 99\% worst-case coupling
  \item Additional new simulations on the validation of the statistical model proposed here are presented.
  \item Additional new simulations are presented to confirm the possibility of using the proposed model for worst-case analysis and for evaluating the effects of raising the PLC transmit PSD in certain bands.

\end{enumerate}

The paper is organized as follows. In Sect. \ref{sec.MeasurementSetUp}, the measurement campaign is described in detail. Considerations on the main characteristics of PLC-to-DSL interference are given in Sect. \ref{sec.ConsiderationsMeasurements}. Statistical considerations on the estimated couplings are given in Sect. \ref{sec.StatsConsiderations}, and the proposed statistical model is set forth in Sect.\ref{sec.TheModel}. Finally, the impact of PLC-to-DSL interference on VDSL2, V-VDSL2, and G.fast is assessed on the basis of both estimated and simulated couplings in the following Sections: in terms of data rate decrease in Sect. \ref{sec.ImpactSims}, and in terms of varying the PLC transmit power above $30$ MHz in Sect. \ref{sec.ImpactRaisePSD}. Conclusive remarks are given in Sect. \ref{sec.Conclusions}.

\section{The Measurement Campaign} \label{sec.MeasurementSetUp}
A PLC-like signal was injected into a set of in-home power line outlets and the interference that coupled into telephone cables in the same home and at various telephone sockets was digitally sampled and stored. Measurements available to this work were carried out in $32$ residential customer premises in France, including houses and apartments both old and new. Interior walls are about about 10 cm thick and are made of plasterboard or brick and plaster. All measurement were performed from Monday to Friday, between hours 9:00 and 18:00.

The domestic phone networks presented the following topologies: bus, star, bus-star and star-bus. The number of phone sockets per location ranged between 1-5, and the median number of sockets was 2. The number of power outlets per location ranged between 3 and 14, and the median number of outlets was 5. This allowed collection of between 4 and 25 PLC interference traces per home - with a median of 11 traces per home. In all cases, the in-home phone cabling was connected to the outside plant.

In France, the preferred solution to deploy DSL is to use microfilters. This means that the customer can install the CPE anywhere in the premises by using a microfilter for POTS/DSL separation. In other words, for the tested configurations, there is no master telephone jack that is close to where the telephone cable enters the home (NID) and isolates the DSL modem from the rest of the noisy inside wiring. Furthermore, in the case of houses, the NID is located in the garage whereas the DSL modem is in the living room. In other countries, the situation may be different and separation from inside wiring would be beneficial to DSL.

A total of 353 time-domain PLC interference traces were collected, where each measurement represents an interference realization between the PLC-like signal injected into a power line outlet and the resulting interference measured at a phone socket in the same home. Also a total of 64 time-domain ambient noise traces on twisted-pairs were collected in the absence of PLC-like signals present on the power cables, and we will refer to these as ``baseline'' noise traces.

\subsection{The Set-Up} \label{subsec.SetUp}
The measurement set-up is shown in Figure \ref{fig.set-up}. An Arbitrary Waveform Generator (AWG, Tektronix AWG5002) was used as a PLC transmitter. By means of a passive coupler, a PLC signal was continuously injected in the electrical outlet.

The coupled PLC noise on the telephone copper pair in the same home is measured at the phone sockets in the time domain using a digital oscilloscope (Lecroy Waverunner 64Xi). A balun was used to adapt the impedance of the oscilloscope (50 Ohms) to the characteristic impedance of the telephone copper pairs (100 Ohms). The waveform generator and the oscilloscope were powered using the home electrical network but, in order to minimize the influence of the measurement equipment on the noise measurements, they were powered using a filtered power supply and they were also powered from a distant outlet. The sampling frequency both at the transmitter and the receiver was set to 250~MHz.

\insertonefig{0}{\FigWidth}{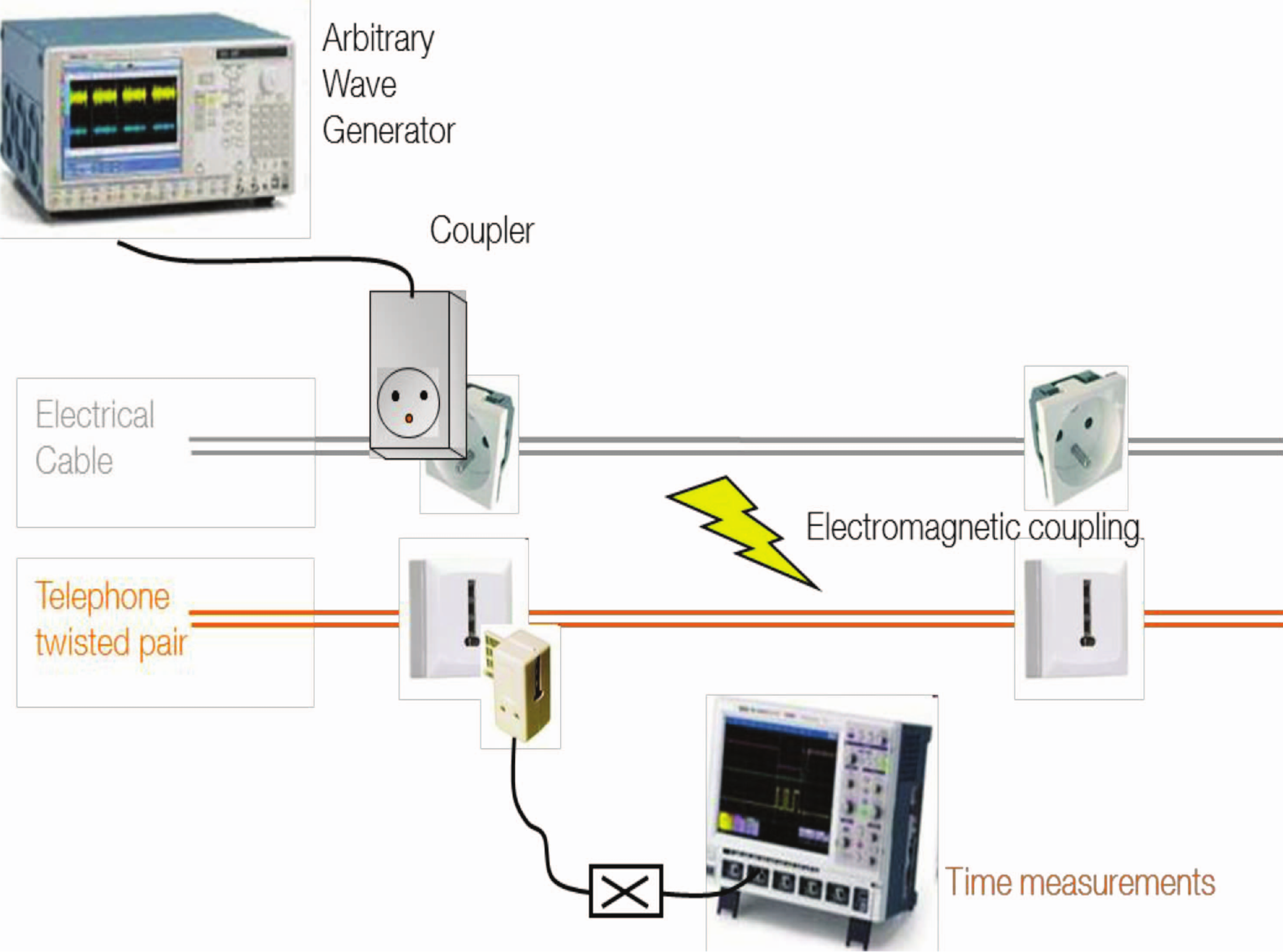}{\label{fig.set-up}}{The measurement set-up.}

\subsection{The injected PLC signal} \label{subsec.EmittedPLC}
The injected PLC-like signal had a flat Power Spectral Density (PSD) of $-60$ dBm/Hz over a band up to 125 MHz. The coupler used for the injection of the PLC noise on the electrical line was custom built, exhibited good wideband characteristics between $0.5-100$ MHz, and introduced only a few dB attenuation. The coupler transfer function is shown in Figure \ref{fig.PLC-Coupler}

\insertonefig{0}{\FigWidth}{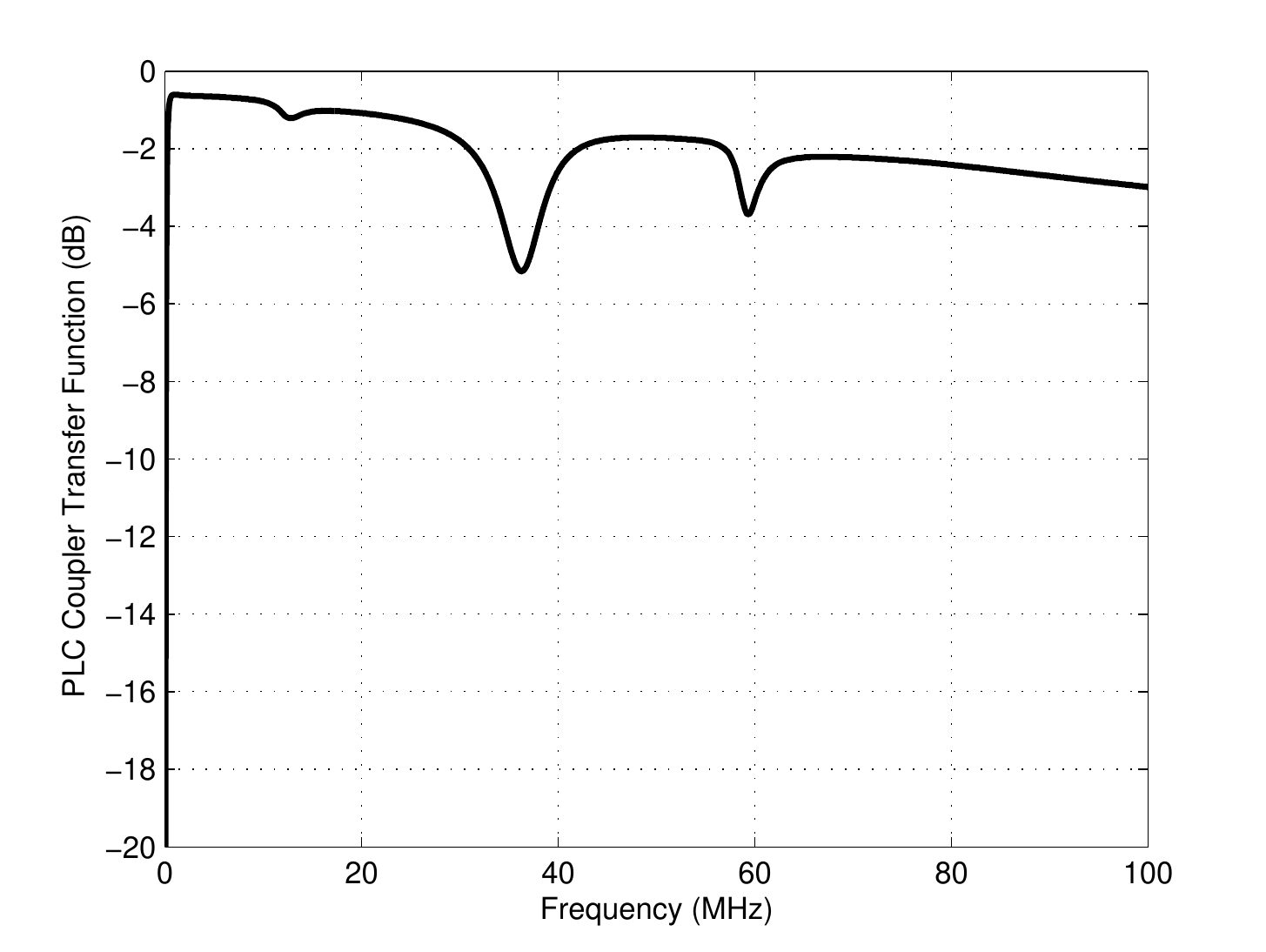}{\label{fig.PLC-Coupler}}{The PLC coupler used in the measurement campaign.}

We point out that both radiated and conducted PLC interference are present on twisted pairs, but in this work we did not attempt to separate the two contributions. We considered the cumulative effect of both components by measuring the interference observed on the twisted pair and by estimating the value of the interference coupling in the frequency domain.

Modern PLC technologies avoid transmitting in the FM broadcast band\footnote{~The FM broadcast band varies in different parts of the world. In ITU Region 1 (Europe and Africa) this band spans $87.5-108$ MHz, while in ITU Region 2 (Americas) it spans $87.9-107.9$ MHz. The FM broadcast band in Japan uses the 76-90 MHz band.} to avoid creating interference to it. ITU-T G.hn allows limiting transmission up 80 MHz and HomePlug AV2 uses frequencies up to 86.13 MHz. The statistical considerations made here for the PLC-to-DSL interference in the $1.8-100$ MHz band apply also to the $1.8-86.13$ MHz band as the estimated couplings Probability Density Function (PDF) for the two bands is basically identical (see Figure 3.(a) in \cite{GalliKerpezMariISPLC2015}).

\subsection{Processing of Measured Interference} \label{subsec.ProcessingTraces}
The baseline and PLC-induced noise present on the customer's telephone copper pairs were measured with a digital oscilloscope sampling at 250 Msamples/sec and connected via an impedance matching balun to a phone socket. Each time-domain trace contained 25 million samples, corresponding to a duration of 100 ms. Each time-domain measured noise trace represents an ``interference realization'' between a power line outlet and a phone socket or simply the ambient baseline noise measured at the phone socket in the absence of PLC-like excitation over the power lines in the same home. There are often multiple traces per home, each from different outlet-phone jack pairs.

The PSD of the interference traces was estimated using a modified Welch algorithm \cite{Harris1978psd}. PLC-to-twisted pair couplings were then computed by subtracting from the estimated noise PSD the PLC-like excitation signal of $-60$ dBm/Hz. The number of estimated couplings available for this study were 1.3 million for the 17 MHz VDSL2 profile VDSL2-17a (1.8-17.6 MHz band), 2.8 million for the 35 MHz VDSL2 profile VDSL2-35b (1.8-35.3 MHz band), and around 680 thousand for the 106 MHz G.fast profile FAST-106 (1.8-100 MHz band).

The estimated interference couplings between power lines and phone lines have been used to simulate PLC-to-DSL interference and assess its impact on the performance of VDSL2, V-VDSL2, and G.fast.

\section{Initial considerations on the Measurements} \label{sec.ConsiderationsMeasurements}
It is today well known that the PLC frequency transfer function is a Linear and Periodically Time-Varying (LPTV) channel \cite{CanCorDie2006}. Recently, it has also been pointed out that the LPTV PLC frequency transfer function is amenable to being represented with a Zadeh decomposition \cite{Galli2010TwoTaps, GallScagXXlptv, GianPancVitetta2014}. Since the PLC frequency transfer function is LPTV and since electro-magnetic coupling can be modeled as a linear transformation, then one would expect that also PLC-to-DSL interference is also LPTV and with the same period. This conjecture has been verified by our measurements.

Figure \ref{fig.Spectrograms} shows two spectrograms from in-home twisted-pairs calculated when (a) PLC interference is present on the twisted-pair, and (b) when it is not. It is easy to notice that, although most frequency components of the channel are static over time, there are clearly some frequency components that change every 10 ms (half of the mains period of 1/50 Hz), switching between two values. The spectrogram in Figure \ref{fig.Spectrograms}.(a) confirms for the first time that the induced PLC-to-DSL interference has a distinct LPTV nature. Furthermore, the spectrogram in Figure \ref{fig.Spectrograms}.(b) confirms that also the baseline noise present on twisted-pairs is LPTV. This observed LPTV noise is due to household appliances and power supplies in the home.

\inserttwofigsV {0}{\FigWidth}{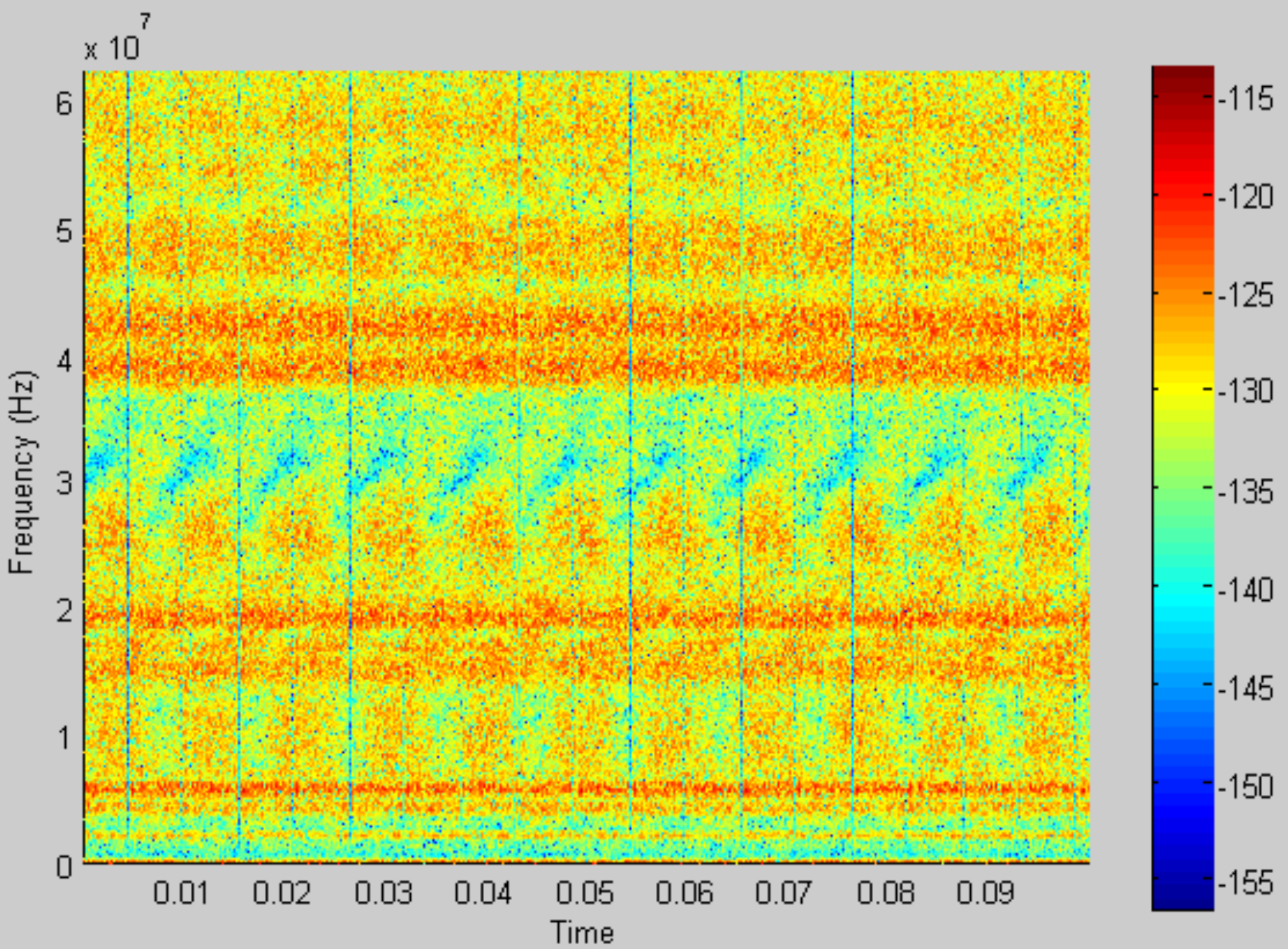}{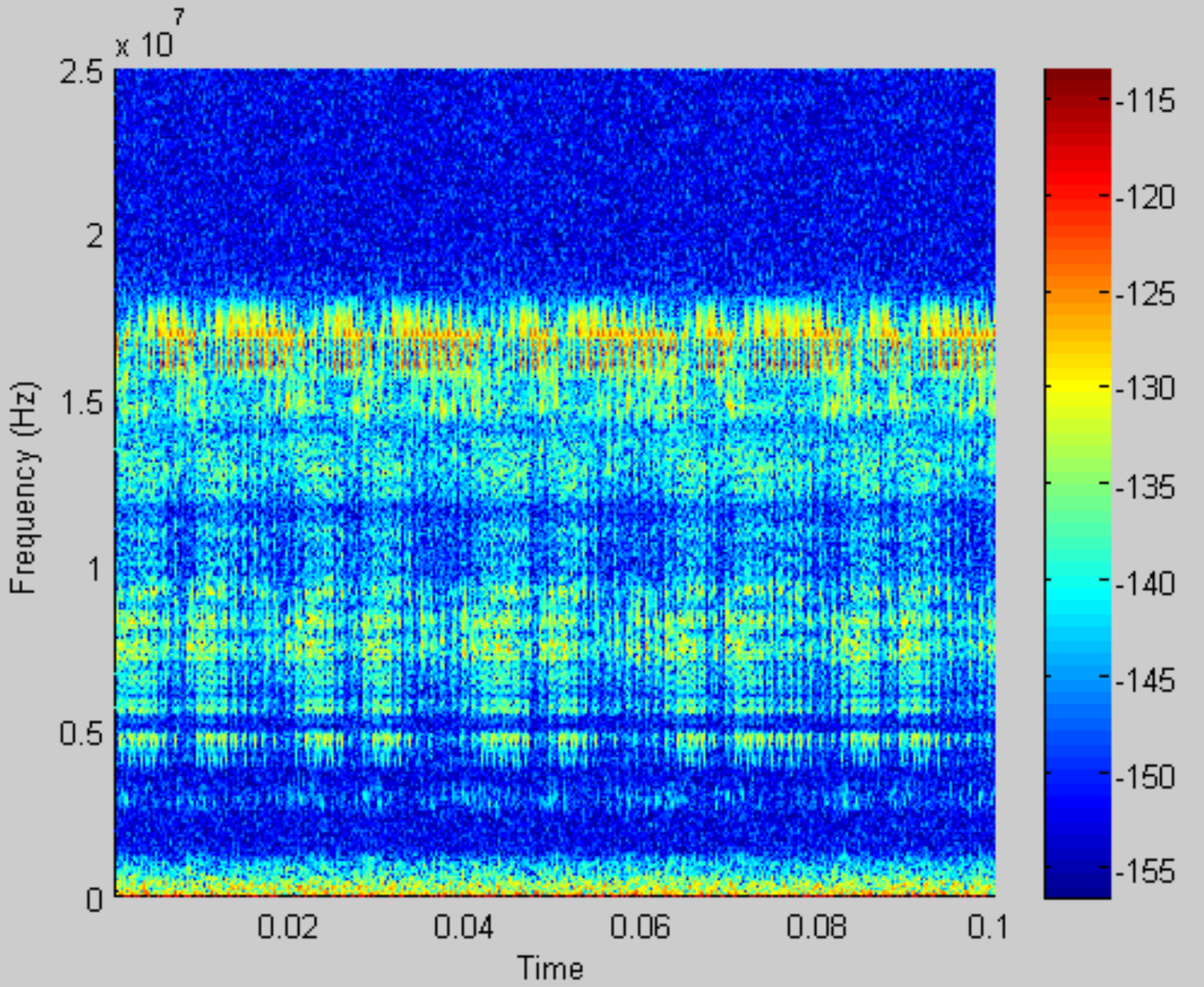}{\label{fig.Spectrograms}}
{Spectrogram of noise measured on in-home phone lines: (a) Active PLC interference trace; (b) Baseline noise trace. Time is in seconds (from 0 to 0.1 s) and frequency is in Hertz.}

Given the LPTV nature of PLC interference, the frequency components of the estimated PSD assume a value that is the time average of the spectrogram along the time-axis. This average PSD is used to extract the PLC-to-DSL interference couplings. In Figure \ref{fig.PSD-AHM}, we show the estimated PSDs of all 64 measured baseline traces and one measured PLC-to-DSL interference trace observed on a telephone jack. It can be verified that the noise floor is sometimes $-150$ dBm/Hz and sometimes $-140$ dBm/Hz.

\insertonefig {0}{\FigWidth}{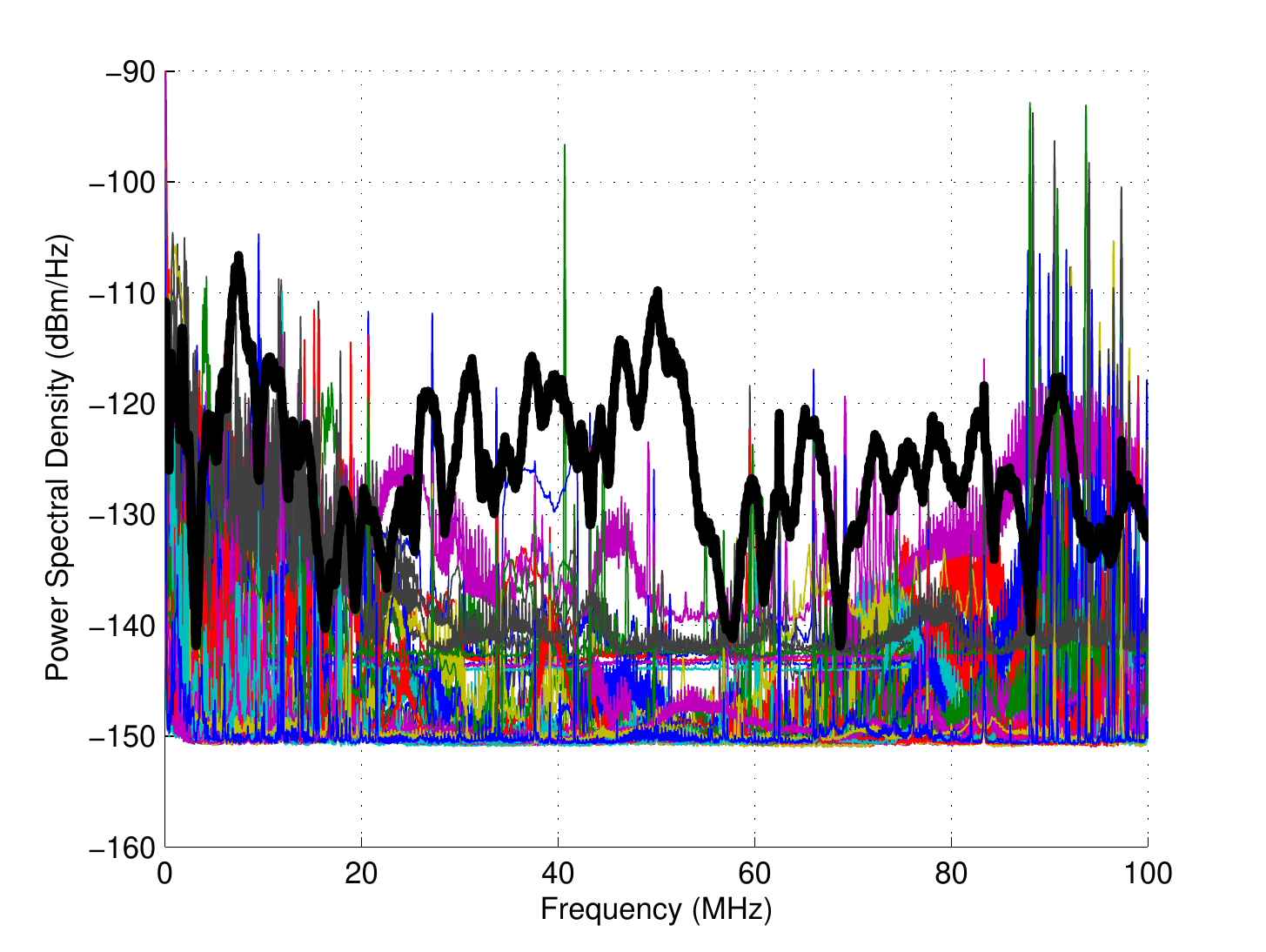}{\label{fig.PSD-AHM}}
{The ambient baseline noise PSDs of all 64 baseline measurements and one measured PLC-to-DSL interference trace (bold black).}

Narrowband interferers (NBIs) are clearly visible, especially in the first 30 MHz of the band (shortwave radio) and in the $87-100$ MHz band (FM radio broadcast). In this work, we made no attempt to isolate NBIs so some estimated couplings may be higher than they would have been in the absence of NBIs. Similarly, other noises are also present on the telephone lines even when PLC is off and this noise can sometimes be substantial and as high as the baseline noise, see Figure \ref{fig.PSD-AHM}. Since the estimated PLC-to-DSL couplings can never fall below the measured ambient noise, which changes from location to location, then some PLC-to-DSL coupling estimates may be higher than their true value and this happens when the true coupling level is below the ambient noise level.

Couplings have high variability across frequency and this variability can be very large, e.g. up to around 50 dB over the noise floor. Some worst-case percentiles of all the $353$ interference realizations are shown in Figure \ref{fig.WCcouplings} for the FAST-106 case. As opposed to the case of crosstalk 99\% worst-case power that grows with frequency \cite{GalKer02}, PLC-to-DSL interference does not appear to grow with frequency and actually appears to be somewhat flat across the whole band with a variability contained within 10 dB. This behavior is different from what has been reported about crosstalk between twisted-pairs, where crosstalk increases with frequency \cite{GalKer02}.

\insertonefig {0}{\FigWidth}{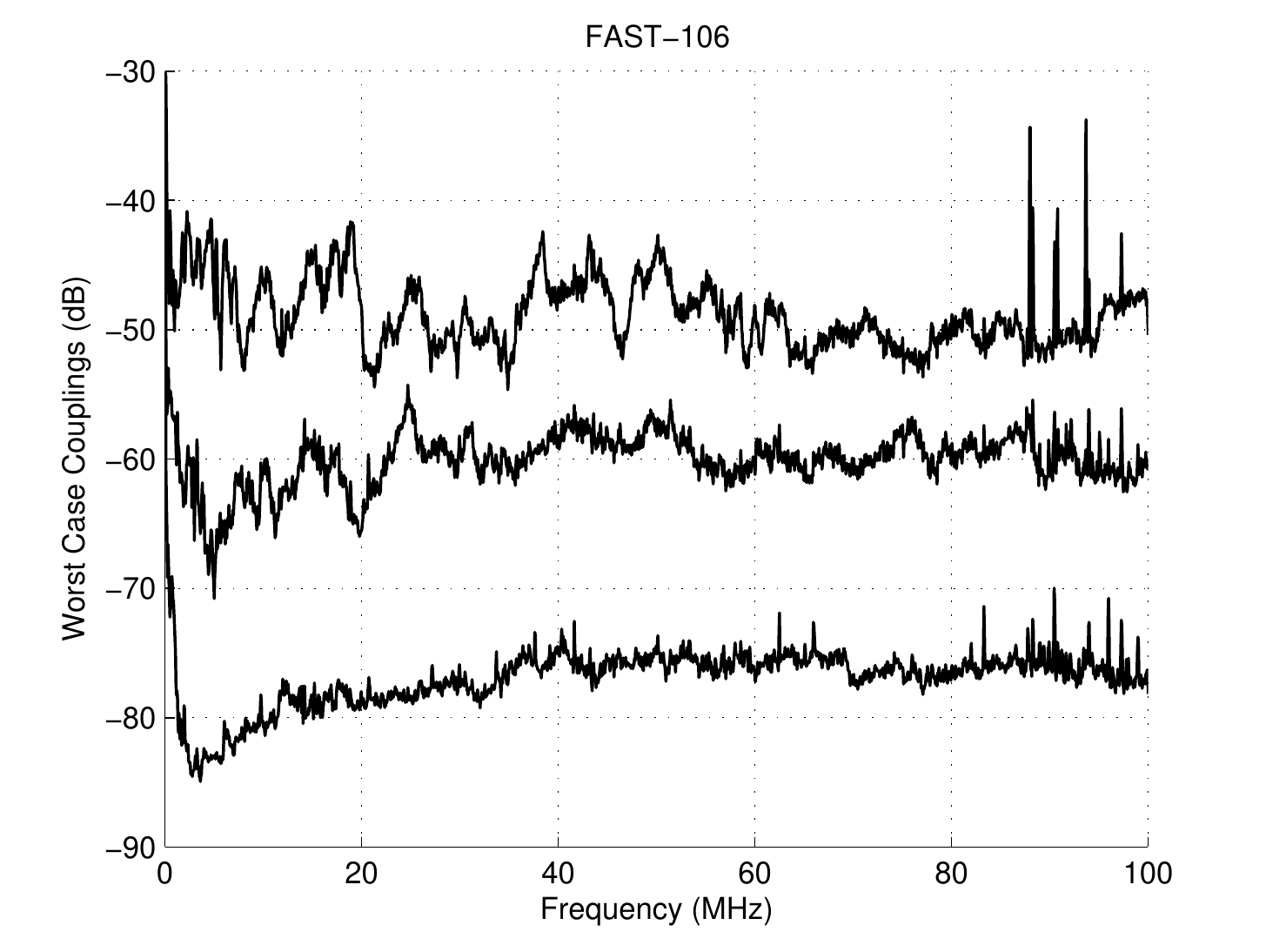}{\label{fig.WCcouplings}}
{Some typical worst-case estimated couplings for the FAST-106 case. From top to bottom: 99\%, 90\%, and 50\%.}

\section{Statistical Considerations on the Couplings Estimated from Measurements} \label{sec.StatsConsiderations}
Table \ref{Tab.PLCint-stats} reports notable statistics for the PLC-to-DSL interference couplings in dB for the case of FAST-106. We remark the large excursion of couplings, nearly 60 dB. Excess Kurtosis, i.e. classical Pearson's kurtosis minus 3, is close to zero, thus showing that the couplings distribution is neither more nor less outlier prone than a Gaussian distribution. Interestingly, there is almost the same difference (around $2\sigma$) between the median and the 90\% worst-case and between the 90\% and 99\% worst-cases.

\begin{table}[htbp]
  \centering
  \caption{Notable statistics of PLC-to-DSL interference couplings in dB in the 1.8-100 MHz band.}
  \label{Tab.PLCint-stats}
\begin{tabular}{l S[table-format=2.1]}
\toprule {} & {Couplings (dB)} \\
\midrule
Min                       	&	-91.0	\\
Max                      	&	-33.6	\\
Mean                     	&	-75.2	\\
Std. Dev. (unbiased)       	&	10.5	\\
Kurtosis                    &	2.9	   \\
Skewness                    &	0.7   \\
50\%-Percentile          	&	-77.0	\\
90\%-Percentile          	&	-59.8	\\
99\%-Percentile             &	-47.8	\\
\bottomrule
\end{tabular}
\end{table}

In the next Sections, we will discuss the estimated Probability Density Function (PDF) of couplings in dB and their Quantile-Quantile (QQ) plots.

\subsection{The Estimated Probability Density Functions} \label{subsec.EstimatedPDF}
The Probability Density Functions (PDFs) of the estimated couplings lying on the frequency grid of VDSL2-17a, VDSL2-35b, and FAST-106 are shown in Figure \ref{fig.PDFestimate}, both in linear (a) and log (b) scales. The PDFs have been estimated via Gaussian kernel smoothing \cite{KernelSmoothing:1994} and the population includes all couplings regardless of frequency or location.

The choice of putting in the same population all estimated couplings will lead to a statistical model where couplings are assigned to carriers randomly and independently. The advantage of this approach is that a single model can be set forth rather than having multiple models devised on a per-frequency or per-band basis. Although this method is able to reproduce (over many realizations) the statistics of estimated couplings, it does not emulate well a single realization of interference between a specific outlet and phone jack as the information on the correlation between couplings across frequency would be lost. However, as simulations in Sect. \ref{sec.ImpactSims} will confirm, the loss of this information will not prevent us from getting accurate results on the average DSL data rate in the presence of PLC interference, as well as accurate worst-case (outage) analysis based on worst-case couplings. In the case it is important to replicate the correlation between couplings across frequencies that is typical of a single realization, this can be done by generating fewer random couplings than the number of DSL sub-carriers, spacing them across the band of interest, and then interpolating the couplings for example via polynomial fit or cubic spline.

\inserttwofigsV {0}{\FigWidth}{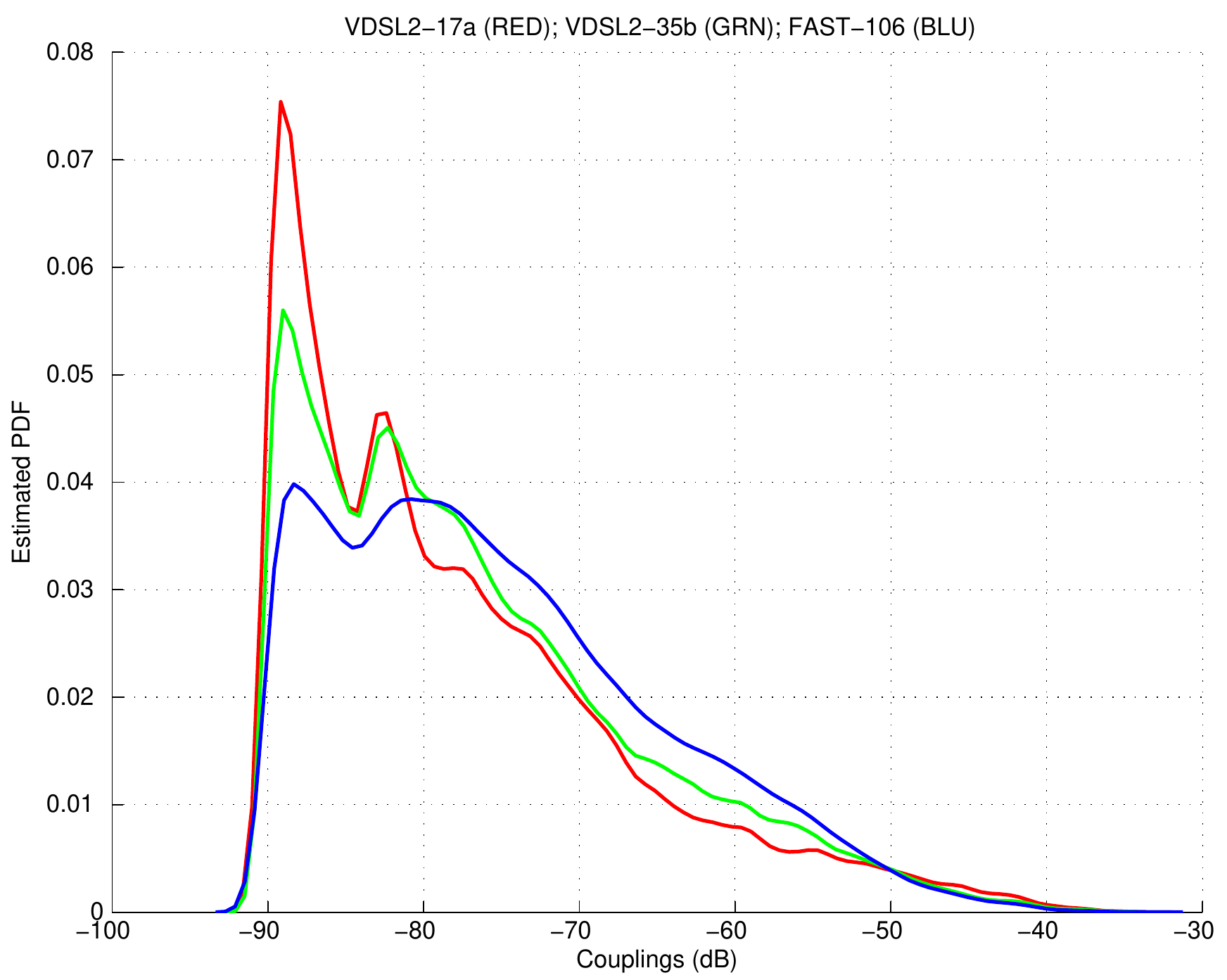}{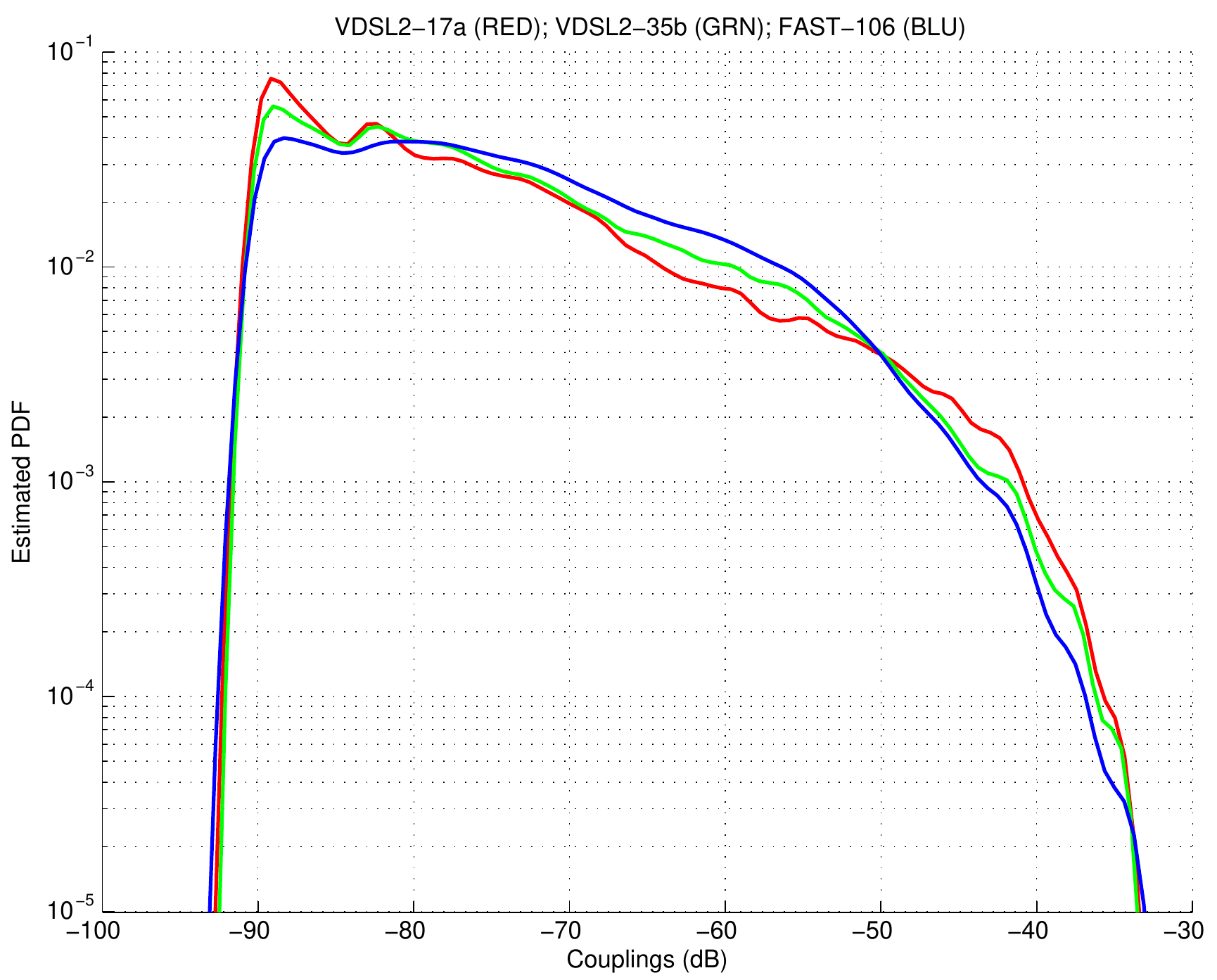}{\label{fig.PDFestimate}}
{Gaussian kernel smoothed estimate of the PDF of the couplings lying on the frequency grid of VDSL2 Profile 17a (red), Profile 35b (green), and G.fast 106 MHz (blue).}

The effect of the noise floor at $-150$ dBm/Hz is noticeable by looking at truncation of the estimated PDF around $-90$ dB. However, the left side of the estimated PDF in Figure \ref{fig.PDFestimate}.(a) seems to suggest that there is a large number of low couplings in the range between $-83$ dB and $-90$ dB. This behavior could be due to the fact that the values of the estimated couplings are never below the baseline noise and, therefore, couplings with low values can be estimated to have higher values than what they should have had if no baseline noise were present.

Two interesting observations about the PDFs in the three DSL bandplans are can be made:
\begin{enumerate}
  \item For the VDSL2-17a case, both very low values and very high values of couplings are much more likely to occur compared to the VDSL2-35b and FAST-106 cases whereas intermediate values of couplings are less likely to occur.
  \item The PDF for the VDSL2-35b case lies somewhat in-between the two PDFs of VDSL2-17a and FAST-106.
\end{enumerate}

\subsection{QQ Plots} \label{subsec.QQplots}
The quantile-quantile (QQ) plots of the couplings lying on the frequency grids of VDSL2-17a and FAST-106 are shown in Figure \ref{fig.QQplot}. The case of VDSL2-35b is intermediate to the ones reported in Figure \ref{fig.QQplot} but it is not reported for the sake of brevity. Apart from the obvious evidence of truncation, the QQ-plots exhibit a rather linear trend for low to moderate couplings and then start deviating above the reference line. This type of QQ-plot may be interpreted as suggesting the presence of mixture of PDFs: a Gaussian one for low-to-moderate couplings and a lighter-tailed distribution for moderate-to-high couplings, e.g. a second Gaussian distribution with a variance smaller than the variance of the first Gaussian.

\inserttwofigsV {0}{\FigWidth}{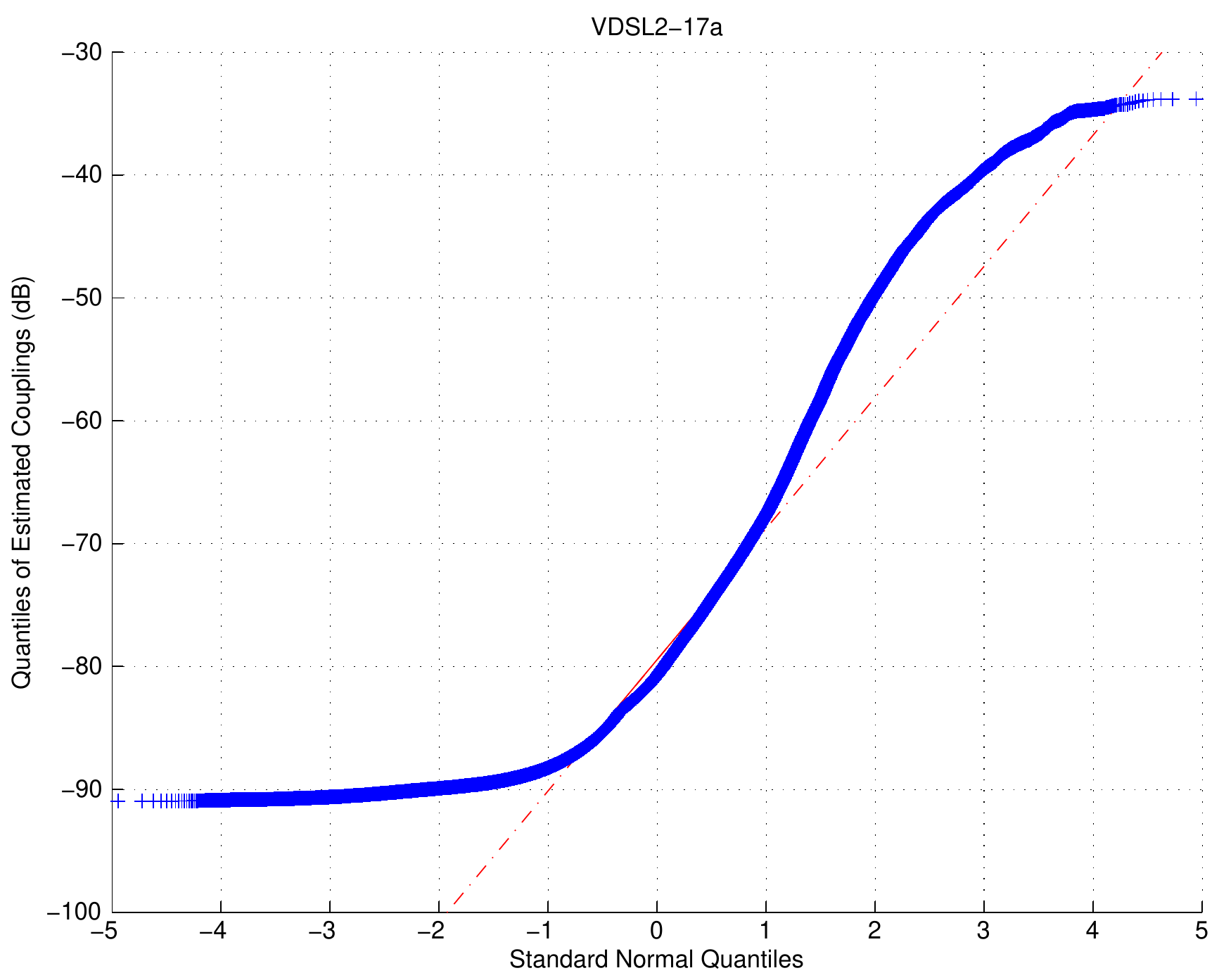}{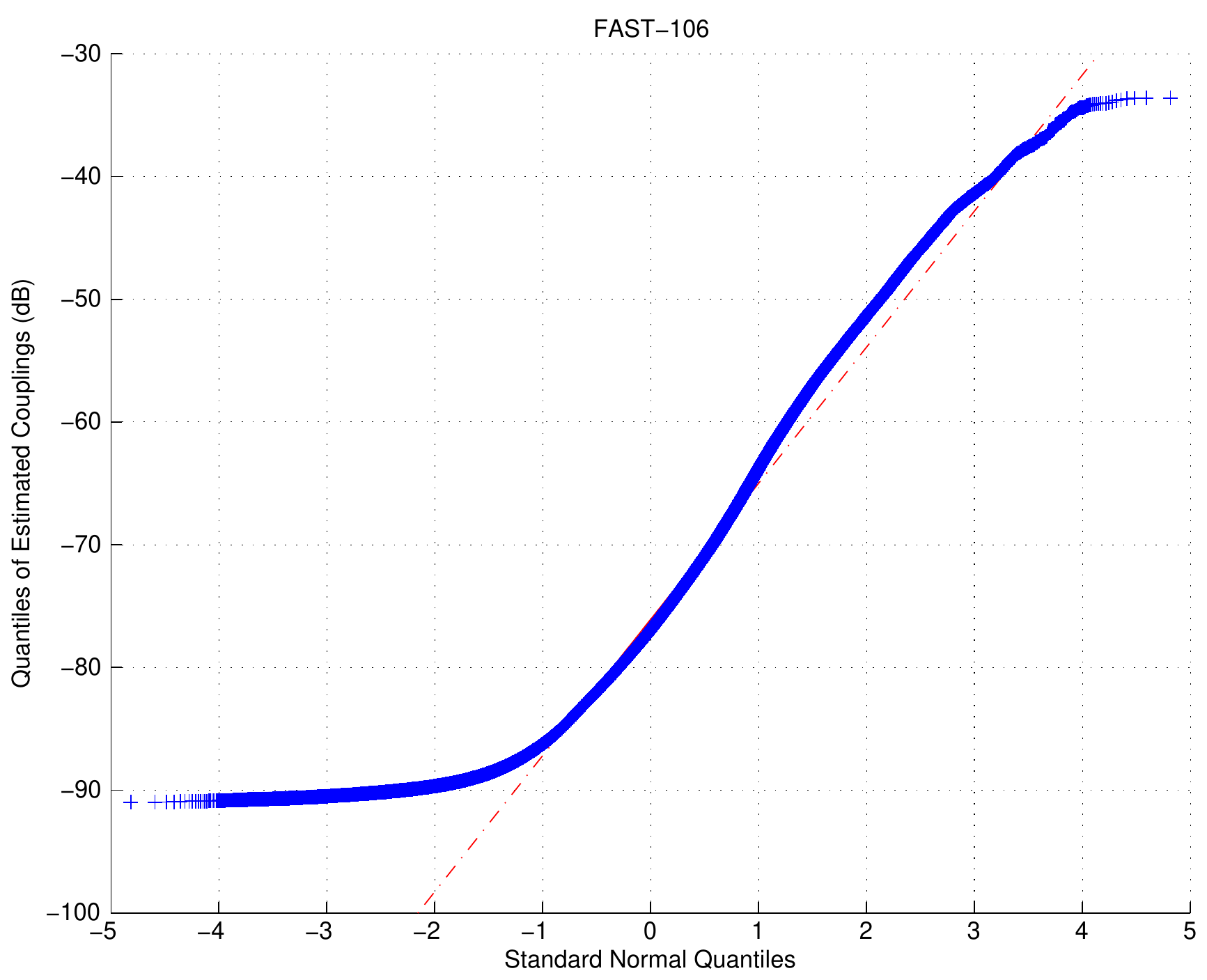}{\label{fig.QQplot}}
{QQ-plot of estimated couplings: (a) VDSL2-17a; (b) FAST-106.}

\section{The Proposed Statistical Model} \label{sec.TheModel}
We have ascertained that the empirical distribution of PLC-to-DSL interference couplings can be well fitted by two truncated Gaussian distributions plus a mass PDF that accounts for the weak couplings below a certain threshold $LT_1$ in all three considered cases (VDSL2-17a, VDSL2-35b, and FAST-106). We have also ascertained that a single fit for both the VDSL2-35b and FAST-106 cases yields to accurate results.

By considering the estimated PDFs, the QQ-plots, and the fitting results presented in this Section, we conclude that there is no strong evidence against modeling PLC-to-DSL couplings as a mixture of two truncated Gaussian distributions. We point out that we have not resorted here to statistical tests for confirming the nature of the distribution of the couplings via $p$-value analysis, as for example done in previous work \cite{Galli2009lognormal, Galli2010TwoTaps, Galli2011wirelinemodel}. The reason for this is that it is well known that the $p$-value inevitably decreases towards zero as sample size increases \cite{Chatfield:95}. For the large sample size we had at our disposal, the $p$-values associated to most statistical tests against the null hypothesis would have been always very close to zero, thus making such investigation moot. On the other hand, the abundance of estimated couplings allows us to rely well on probability plots, histogram analysis, and fitting results.

The following parametric model for the PDF of the couplings $x$ when expressed in dB can be set forth as follows:
\begin{IEEEeqnarray}{rCl}
p(x) & = & Pr\{X \leq LT_1\}\delta(x-LT_1) + \nonumber\\
    && +\: Pr\{X \in \cbL=[LT_1,RT_1]\}\cbT\cbN^{(1)}_{[LT_1,RT_1]}(\mu_1,\sigma_1) + \nonumber\\
    && +\: Pr\{X \in \cbM=[LT_2,RT_2]\}\cbT\cbN^{(2)}_{[LT_2,RT_2]}(\mu_2,\sigma_2), \nonumber\\*
\label{eq.PDFmodel}
\end{IEEEeqnarray}

\noi where $\cbT\cbN^{(i)}_{[a,b]}(\mu_i,\sigma_i)$ denotes a Gaussian distribution with mean $\mu_i$ and standard deviation $\sigma_i$ truncated to the interval $[a,b]$, and $\delta(x)$ is the Dirac delta function. If we denote as $\phi(x)$ the standard normal PDF with zero mean and unitary standard deviation and with $\Phi(x)$ its Cumulative Distribution Function (CDF), then we can express the truncated Gaussian PDFs in eq. \eqref{eq.PDFmodel} as follows ($i=1,2$):

\begin{equation}\label{eq.TruncNormal}
    \cbT\cbN^{(i)}_{[a,b]}(\mu_i,\sigma_i) = \frac{\frac{1}{\sigma_i}\phi(\frac{x-\mu_i}{\sigma_i})}{\Phi(\frac{b-\mu_i}{\sigma_i})-\Phi(\frac{a-\mu_i}{\sigma_i})},
\end{equation}

The fitting of the empirical PDFs in intervals $\cbL$ and $\cbM$ with two truncated Gaussians was carried out using the Levenberg-Marquardt algorithm \cite{BatesWatts:88}. This algorithm directly solves non-linear regression problems thus averting the usual pitfalls associated with linear regression methods that resort to exponential transformations and thus tend to overly-weight outliers. The probabilities in eq. \eqref{eq.PDFmodel} have been obtained from the Kaplan-Meier estimate of the empirical Cumulative Distribution Function (CDF) of the measurements \cite{KapMei1958}.

Besides being statistically valid, this approach is also physically meaningful as PLC and DSL signal propagation gives rise to normally distributed fading (in dB) \cite{Galli2009lognormal, Galli2010TwoTaps, Galli2011wirelinemodel}, and this distribution should be preserved by electromagnetic coupling as coupling can be viewed as a form of filtering (linear transformation).

The parameters of the fitting mixture PDF in eq. \eqref{eq.PDFmodel} depend on the considered bandplan so these will addressed separately.

\subsection{The VDSL2-17a Case} \label{subsec.Model-17a}
The parameters of the best (in the Minimum Squared Error sense) fitting mixture PDF for the VDSL2-17a case are:

\begin{itemize}
  \item $Pr\{X \leq -90\}=0.0171$
  \item $Pr\{X \in \cbL=(-90, -44)\}=0.9757$
  \item $Pr\{X \in \cbM=[-44, -33]\}=0.0072$,
\end{itemize}

\noi while the mean and standard deviation of the truncated Gaussian PDFs appearing in eq. \eqref{eq.PDFmodel} are:
\begin{itemize}
  \item $\mu_1=-110.8$ dB, $\sigma_1=24.2$ dB
  \item $\mu_2=-44.6$ dB, $\sigma_2=3.5$ dB.
\end{itemize}%

Figure \ref{fig.Fitted17} shows the estimated and fitted PDFs. We also report in Figure \ref{fig.QQsim17} the QQ-plot of the measured couplings versus the simulated ones. It can be verified that the plot is very linear confirming that the data generated according to the proposed statistical model and the empirical data come from the same distribution.

\insertonefig {0}{\FigWidth}{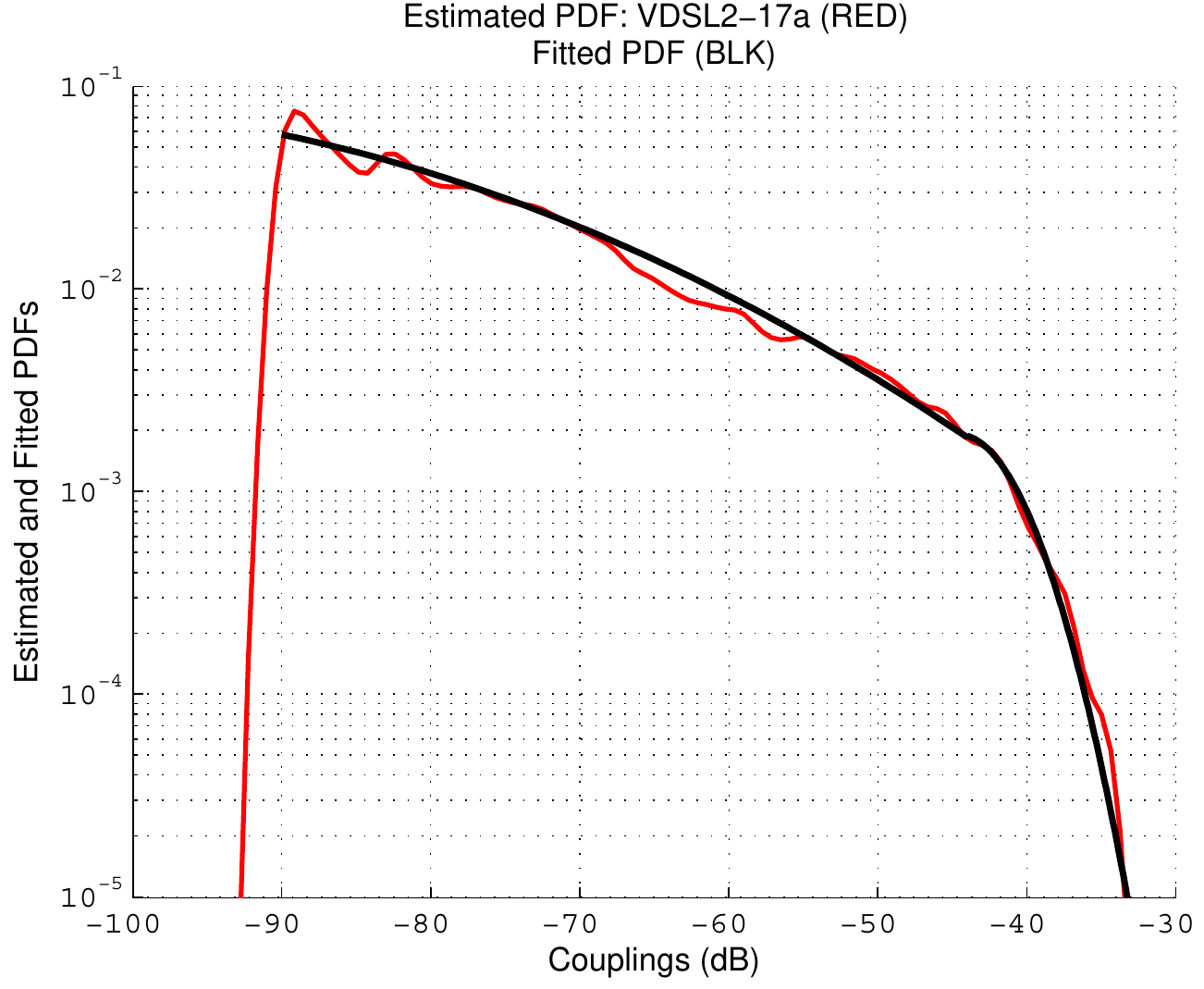}{\label{fig.Fitted17}}
{Plot of the estimated (Gaussian kernel smoothing) and fitted PDFs for the VDSL2-17a bandplan: estimated (red) and fitted (black).}

\insertonefig {0}{\FigWidth}{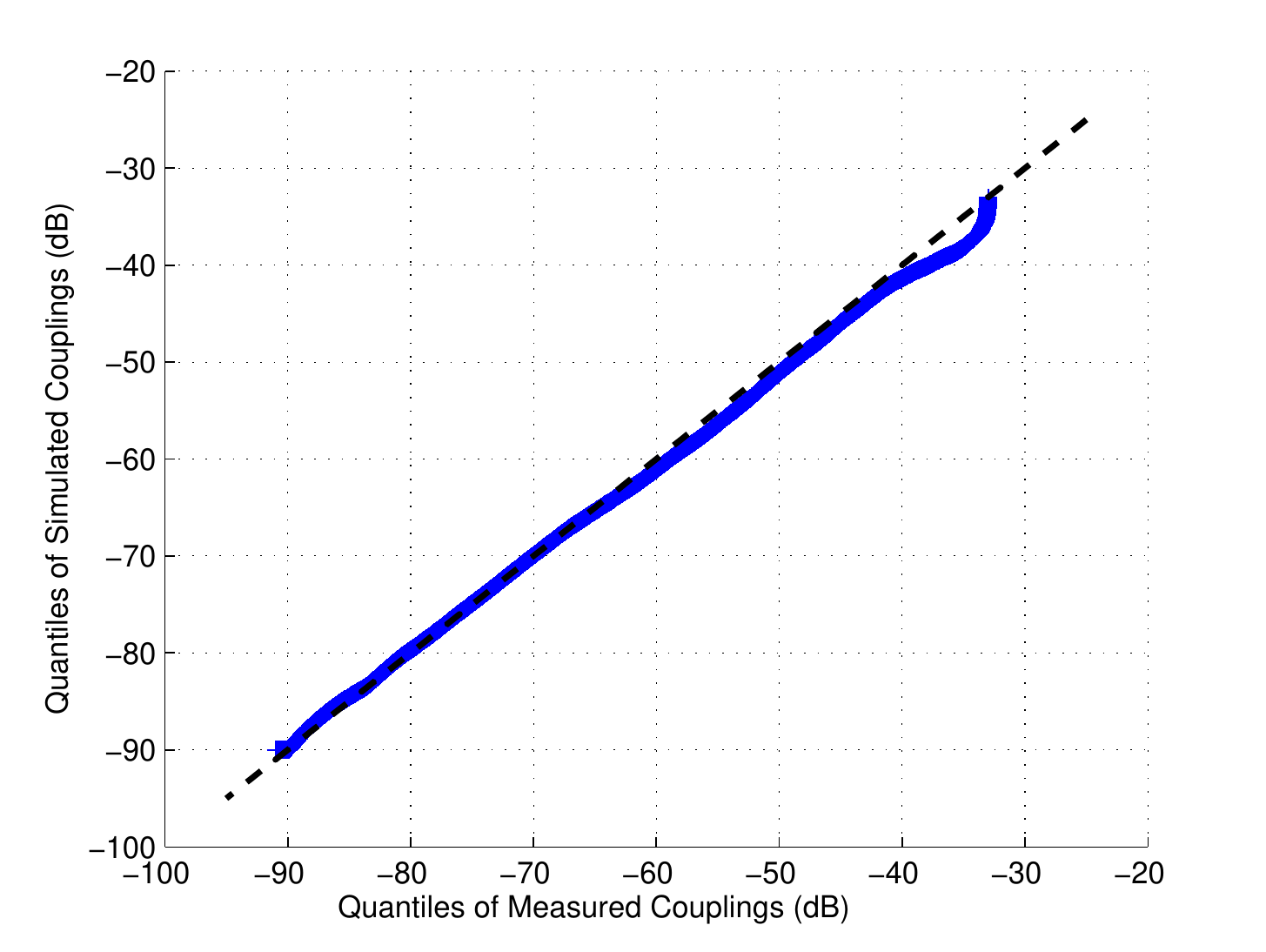}{\label{fig.QQsim17}}
{QQ plot of measured versus simulated couplings. The empirical data has been truncated to -33 dB, the same truncation of the simulated data.}

\subsection{The VDSL2-35b and FAST-106 Cases} \label{subsec.Model-35bFast}
A single fitting distribution can be used for both the VDSL2-35b and FAST-106 cases. The parameters of the best (in the Minimum Squared Error sense) fitting mixture PDF are:

\begin{itemize}
  \item $Pr\{X \leq -90\}=0.0111$
  \item $Pr\{X \in \cbL=(-90, -50)\}=0.9722$
  \item $Pr\{X \in \cbM=[-50, -33]\}=0.0167$,
\end{itemize}

\noi while the mean and standard deviation of the truncated Gaussian PDFs appearing in eq. \eqref{eq.PDFmodel} are:
\begin{itemize}
  \item $\mu_1=-84.0$ dB, $\sigma_1=15.8$ dB
  \item $\mu_2=-60.5$ dB, $\sigma_2=8.1$ dB.
\end{itemize}%

Figure \ref{fig.Fitted35fast} shows the estimated and fitted PDFs. We also report in Figure \ref{fig.QQsim106} the QQ-plot of the measured couplings (both VDSL2-35b and FAST-106 cases) versus the simulated ones according to the single model proposed in this Section. It can be verified that the plot for the FAST-106 is very linear confirming that the data generated according to the proposed statistical model and the empirical data come from the same distribution. The plot for the measured VDSL2-35b data versus the same simulated data show only a slight deviation from linearity thus confirming that the use of a single model for both bands is indeed justified.

\insertonefig {0}{\FigWidth}{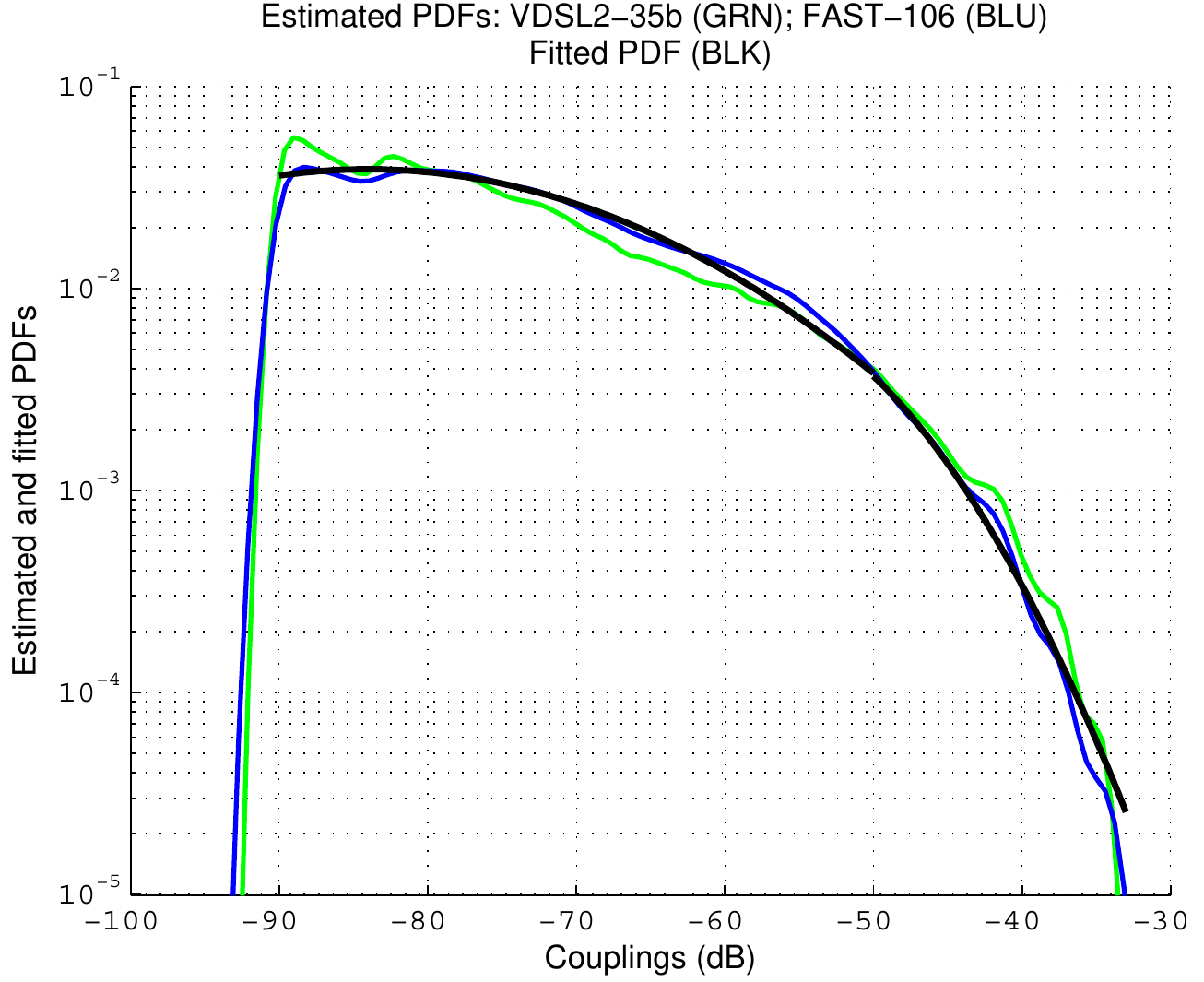}{\label{fig.Fitted35fast}}
{Plot of the estimated (Gaussian kernel smoothing) and fitted PDFs for the VDSL2-35b and FAST-106 bandplans: estimated VDSL2-35b (green), estimated FAST-106 (blue) and fitted (black).}

\insertonefig {0}{\FigWidth}{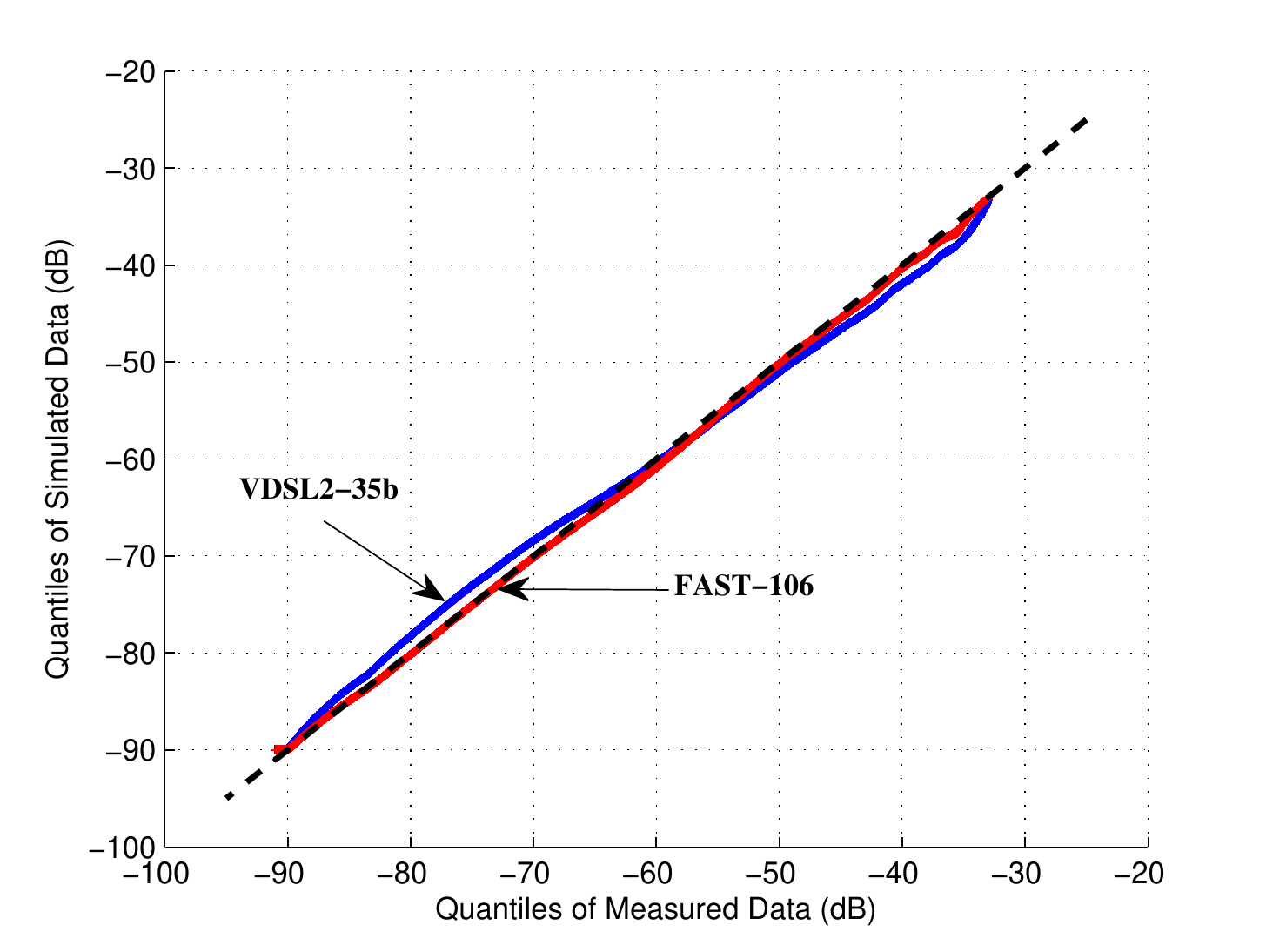}{\label{fig.QQsim106}}
{QQ plot of VDSL2-35b (blue) and FAST-106 (red) measured versus the same set of simulated couplings. The two empirical data sets have been truncated to -33 dB, the same truncation of the simulated data.}

\section{Evaluation of the Impact of PLC-to-DSL Interference on DSL Data Rate}  \label{sec.ImpactSims}
The simulations presented in this section assess the impact of PLC-to-DSL interference using the couplings estimated on the basis of measurements or generated via the statistical model in Sect. \ref{sec.TheModel}. Unless otherwise specified, the data rates when PLC interference is absent have been calculated using the measured baseline traces.

\subsection{Simulation Methodology} \label{subsec.SimsMethod}

\subsubsection{PLC Interference} \label{subsubsec.MethodPLCinterf}
PLC interference was generated using a maximum allowed PSD of -55 dBm/Hz in 2-30 MHz and of -85 dBm/Hz in 30-87 MHz, while no PLC transmission is present in the FM broadcast band of 87-100 MHz. This choice is consistent with the recent HomePlug AV2 \cite{Book:PLC-MIMOandNB-2014} and ITU-T G.hn \cite{OksGal2009} specifications. We point out that most PLC devices in the field today transmit up to 67.5 MHz so that allowing PLC to go up to 87 MHZ in our simulations will translate in a larger impact to DSL.

This transmit PSD is added, in dB, to the estimated couplings to form a realization of the noise PSD into the DSL downstream receiver. This methodology assumes that the PLC node is continuously transmitting at maximum power, which is not the case under normal operations.

\subsubsection{VDSL2 and V-VDSL2} \label{subsubsec.MethodDSL}
VDSL2 uses high frequencies and loop lengths typically up to 1.5~km to transmit at speeds of at most a few hundred Mbps. VDSL2 uses frequency-division duplexing, upstream and downstream, to avoid near-end crosstalk (NEXT). Vectoring removes the FEXT created within a Vectored group by performing precoding at the transmitter (downstream) and crosstalk cancelation at the receiver (upstream). Thus, Vectoring allows maintaining single-line VDSL2 data rate regardless of the number of self-crosstalkers.

The simulations here calculate downstream (V-)VDSL2 bit rates. The simulations use models and parameters from ANSI Std. T1.413 \cite{STD:ANSI:T1.417}. Both VDSL2 Profile 17a (VDSL2-17a) and Profile 35b (VDSL2-35b) are simulated. For VDSL2-17a, the transmit PSD is at most 3.5 dB below the VDSL2 profile 998ADE17-M2x-A PSD limit mask defined in Annex B of G.993.2 \cite{STD:ITU:VDSL2-2015}. For VDSL2-35b, the Draft Annex additionally specifies -60 dBm/Hz from 17.7 to 30 MHz and -76.7 dBm/Hz above 30 MHz, all downstream. Additionally, the maximum transmit PSD is limited to meet the G.993.2 average power constraint \footnote{The maximum PSD level is lowered to a point where the transmit power across all frequencies meets the total power constraint. The same thing is done for G.fast. This is common industry practice.} and there is no Downstream Power Back-Off. AWGN at -140 dBm/Hz is also added in all cases.

For VDSL2, the simulations include two 99\% worst-case FEXT crosstalkers. For V-VDSL2, the simulations model the ideal case of complete in-domain crosstalk cancelation and no out-of-domain/alien crosstalk present \cite{STD:BBF:TR-320}. The measured baseline noise is added to the received signal.

For the downstream data rate calculation, margin is 6 dB and total coding gain is 3 dB. Bit rates are calculated by summing the capacity of each 4.3125 kHz sub-carrier with a 9.75 dB SNR gap, with bits per Hz per sub-carrier lower limited to at least one bit and upper limited to 14 bits/Hz per sub-carrier. A guard-band of 12 sub-carriers is imposed between each passband and these guard-bands carry no bits. Loops are all 0.5mm/24 AWG. VDSL2 has 10\% overhead but all presented data rates are pure line rates.

\subsubsection{G.fast} \label{subsubsec.MethodGfast}
G.fast was simulated for both the 106 MHz and 212 MHz Profiles. For both profiles, the following common assumptions were made:

\begin{itemize}
  \item The transmit frequency band goes from 2.5 MHz up to 106 MHz or 212 MHz.
  \item A full cable of 10 active and Vectored lines was simulated.
  \item The G.fast PSD is always normalized, so that the transmit PSD is unaltered by precoding at all frequencies.
  \item Loop response and FEXT are measured on a 100m 0.5mm/24 AWG distribution cable\footnote{~There is no commonly agreed upon channel model for twisted-pairs that extends beyond 50 MHz. Thus, for G.fast, we limit our analysis to loop lengths of 100 meters only for which there is the availability of measurements performed by British Telecom \cite{ITU:BTmeasuredcables}.}.
  \item Background noise is simulated using the measured baseline noise traces, and AWGN is -140 dBm/Hz below 30 MHz and -150 dBm/Hz above 30 MHz.
\end{itemize}

\noi For the 106 MHz case (FAST-106) we additionally have:

\begin{itemize}
  \item DMT with 2048 carriers equally spaced between DC and 106 MHz, at a spacing of 51.75 kHz.
  \item Vectoring is performed using linear zero-forcing precoders. No channel estimation is performed (perfect knowledge of the channel matrix is assumed at the transmitter) and there is no precoder quantization error.
  \item The G.fast PSD is limited as in Recommendation ITU-T G.9700, with PSD mask of -65 dBm/Hz below 30 MHz, and linearly sloping from -73 dBm/Hz at 30 MHz down to -76 dBm/Hz at 106 MHz. This is further limited to meet the 4.0 dBm total transmit power limit by limiting the max transmit PSD to -76.15 dBm/Hz when transmitting from 2.5 to 106 MHz.
\end{itemize}

\noi For the 212 MHz case (FAST-212), we also have:

\begin{itemize}
  \item DMT with 4096 carriers equally spaced between DC and 212 MHz, at a spacing of 51.75 kHz.
  \item Vectoring is performed using a non-linear precoder based on the QR factorization of the complex conjugate of channel matrix as described in \cite{GinisCioffi:2000}. No channel estimation is performed (perfect knowledge of the channel matrix is assumed at the transmitter) and there is no precoder quantization error.
  \item The PSD is the same as that of FAST-106 below 106 MHz, and then exhibits a linear slope from -76 dBm/Hz at 106 MHz to -79 dBm/Hz at 212 MHz.
\end{itemize}

Data rate is calculated by summing the capacity of each 51.75 kHz G.fast sub-carrier with a 9.75 dB SNR gap, with bits per Hz lower limited to at least one bit per Hz and upper limited to 12 bits per Hz.  Margin was set to 6 dB and total coding gain to 3 dB. Differently from the VDSL2 case, speeds of G.fast are maximum total line rate\footnote{~Unlike previous DSL standards, G.fast uses Time Division Duplexing (TDD) and the asymmetry ratio can be varied. Rather than arbitrarily picking an asymmetry ratio, we used the maximum data rate (upstream plus downstream) as if 100\% of resources were devoted to downstream.}. G.fast has 10\% overhead but all presented data rates are net of this overhead.

Since measured cable responses and FEXT were used, the direct and crosstalk channels are different from line to line and this leads to data rate variation across lines even when no PLC-to-DSL interference is present.

\subsection{Assessing Impact on DSL Using Estimated Couplings}  \label{subsec.ImpactResultsMeasuredCouplings}
The simulations presented in this section assess the impact of PLC interference using the couplings estimated on the basis of measurements. In all cases, the data rates when PLC interference is off have been calculated using the measured baseline noise traces.

\subsubsection{VDSL2} \label{subsubsec.ImpactResultsVDSL2}
In Figure \ref{fig.VDSL2Rate}, the downstream data rates of VDSL2 impaired by two 99\% worst-case FEXT crosstalkers and PLC interference are shown versus loop length for both considered profiles. The median curve confirms that, in at least 50\% of cases, the data rate degradation is negligible up to around 400m-500m, and still small at longer loop lengths. However, in the case of strong couplings, the performance degradation can be substantial.

\insertonefig {0}{\FigWidth}{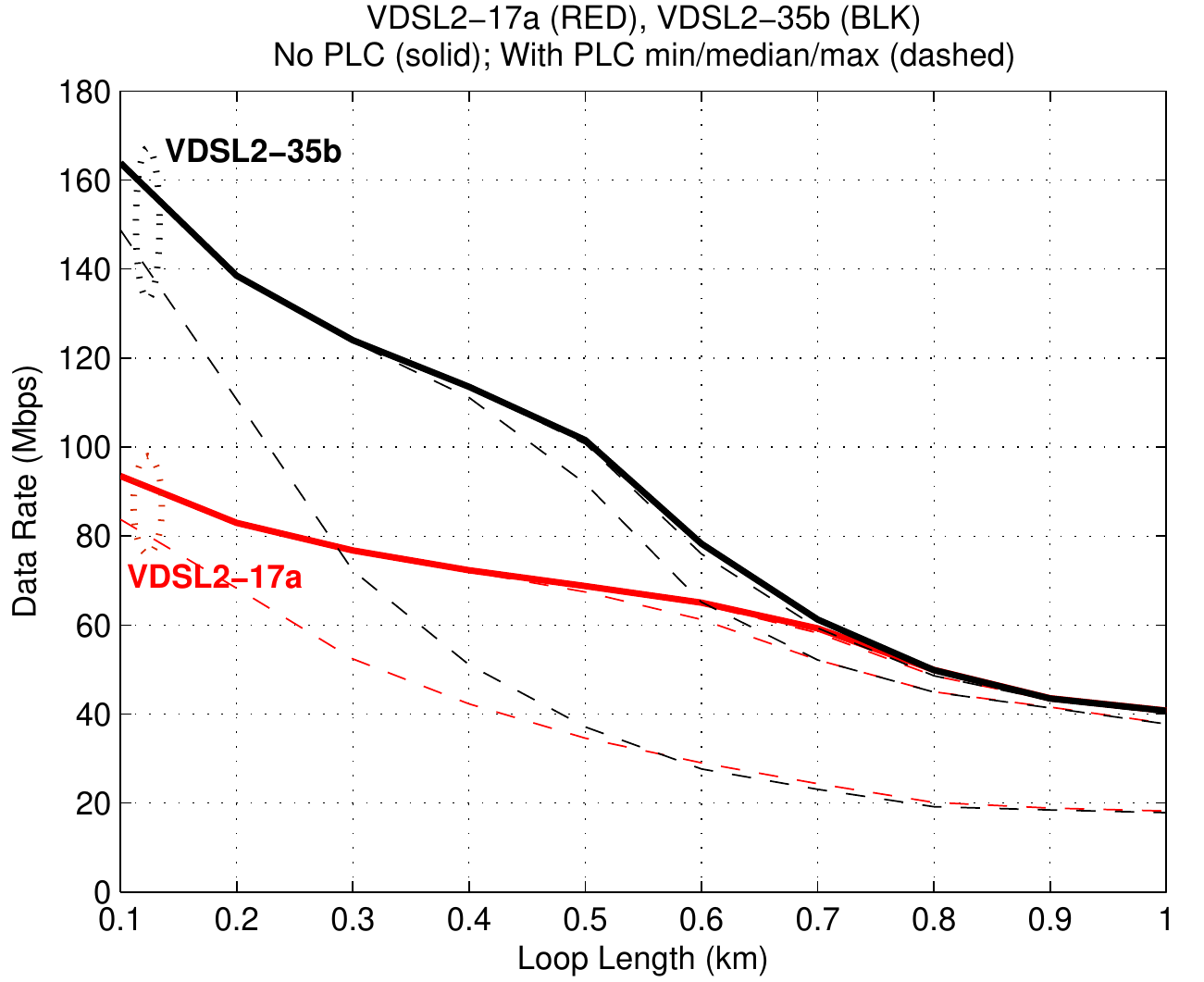}{\label{fig.VDSL2Rate}}
{Data rates achievable by VDSL2 profiles 17a and 35b versus distance, with and without PLC interference. For the performance in the presence of PLC-to-DSL interference, the minimum, median, and maximum data rates are shown as dashed lines. Two 99\% worst-case FEXT crosstalkers are present.}

The average and maximum data rate loss due to PLC interference are plotted in Figure \ref{fig.VDSL2Loss}. As distance grows, the SNR at the DSL receiver decreases due to loop attenuation and the data rate loss increases (percentage-wise). At a certain point, however, the impact on DSL decreases. This is due to two causes: (a) after 500-700 meters the number of active high-frequency sub-carriers decreases due to loop attenuation, while the sub-carriers below 1.8 MHz remain active but are PLC-interference free; (b) there is a higher probability of encountering low coupling values at lower frequencies (see Figure \ref{fig.PDFestimate}).

\insertonefig {0}{\FigWidth}{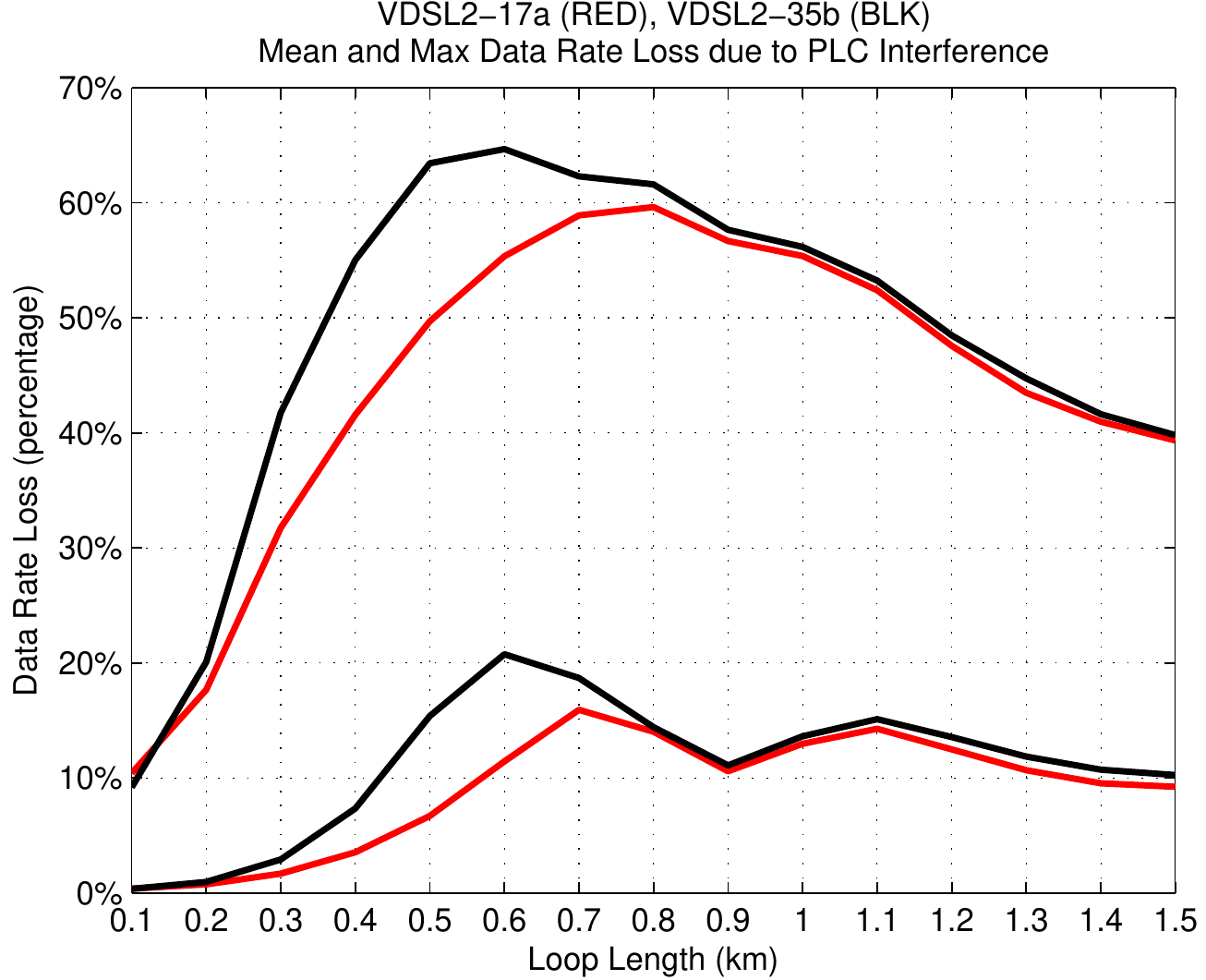}{\label{fig.VDSL2Loss}}
{Mean and maximum data rate loss due to PLC interference for VDSL2 profiles 17a (red) and 35b (black).}

\subsubsection{Vectored VDSL2} \label{subsubsec.ImpactResultsV-VDSL2}
In Figure \ref{fig.V-VDSL2Rate}, the downstream data rates of V-VDSL2 impaired by PLC interference are shown versus loop length for both considered profiles. The median curve confirms that, in at least 50\% of cases, the data rate degradation is limited up to 15\%-25\%. However, in the case of strong couplings, the performance degradation can be substantial.

\insertonefig {0}{\FigWidth}{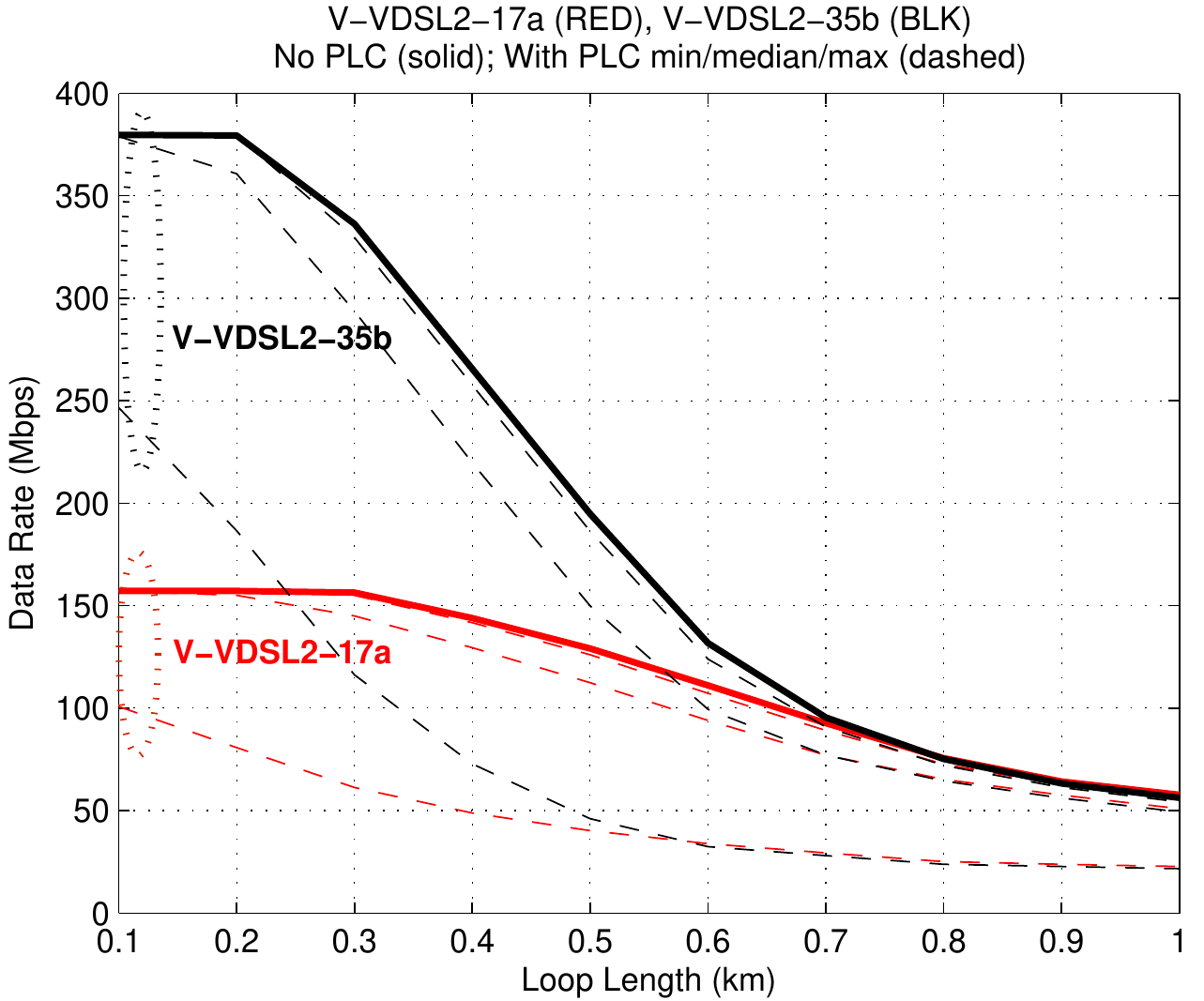}{\label{fig.V-VDSL2Rate}}
{Data rates achievable by V-VDSL2 profiles 17a and 35b versus distance, with and without PLC interference. For the performance in the presence of PLC-to-DSL interference, the minimum, median, and maximum data rates are shown as dashed lines.}

The average and maximum data rate loss due to PLC interference are plotted in Figure \ref{fig.V-VDSL2Loss}. Impact on V-VDSL2 follows the same trend already described for the case of VDSL2.

\insertonefig {0}{\FigWidth}{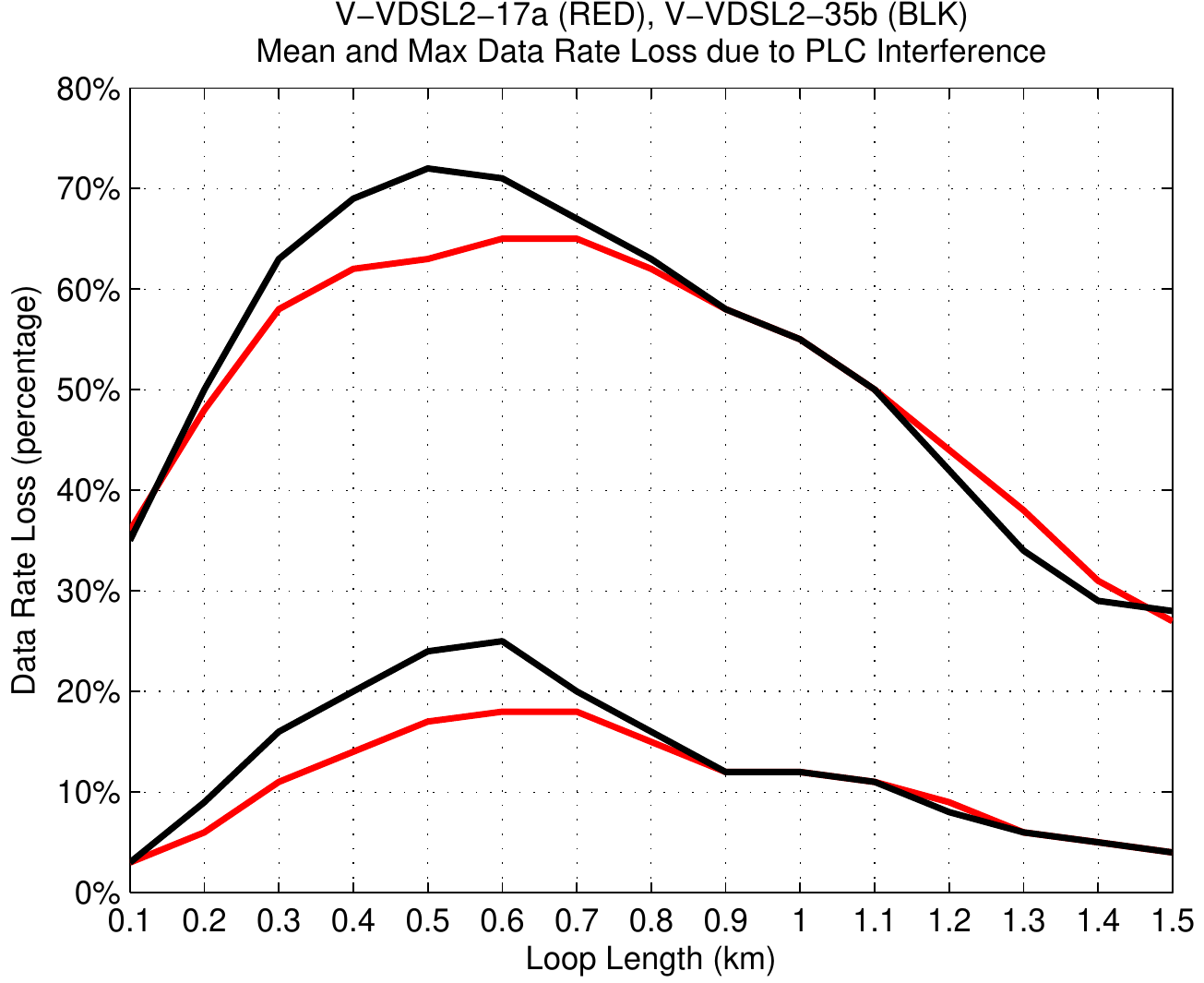}{\label{fig.V-VDSL2Loss}}
{Mean and maximum data rate loss due to PLC interference for V-VDSL2 profiles 17a (red) and 35b (black).}

\subsubsection{G.fast} \label{subsec.ImpactResultsFAST}
The performance of a G.fast system was evaluated on the basis of real loop and FEXT measurements and for a loop length of 100m. The 353 available PLC-to-DSL interference realizations were used for the calculation of the data rates of 10 G.fast Vectored lines, allowing the use of 35 different realizations per line.

\insertonefig {0}{\FigWidth}{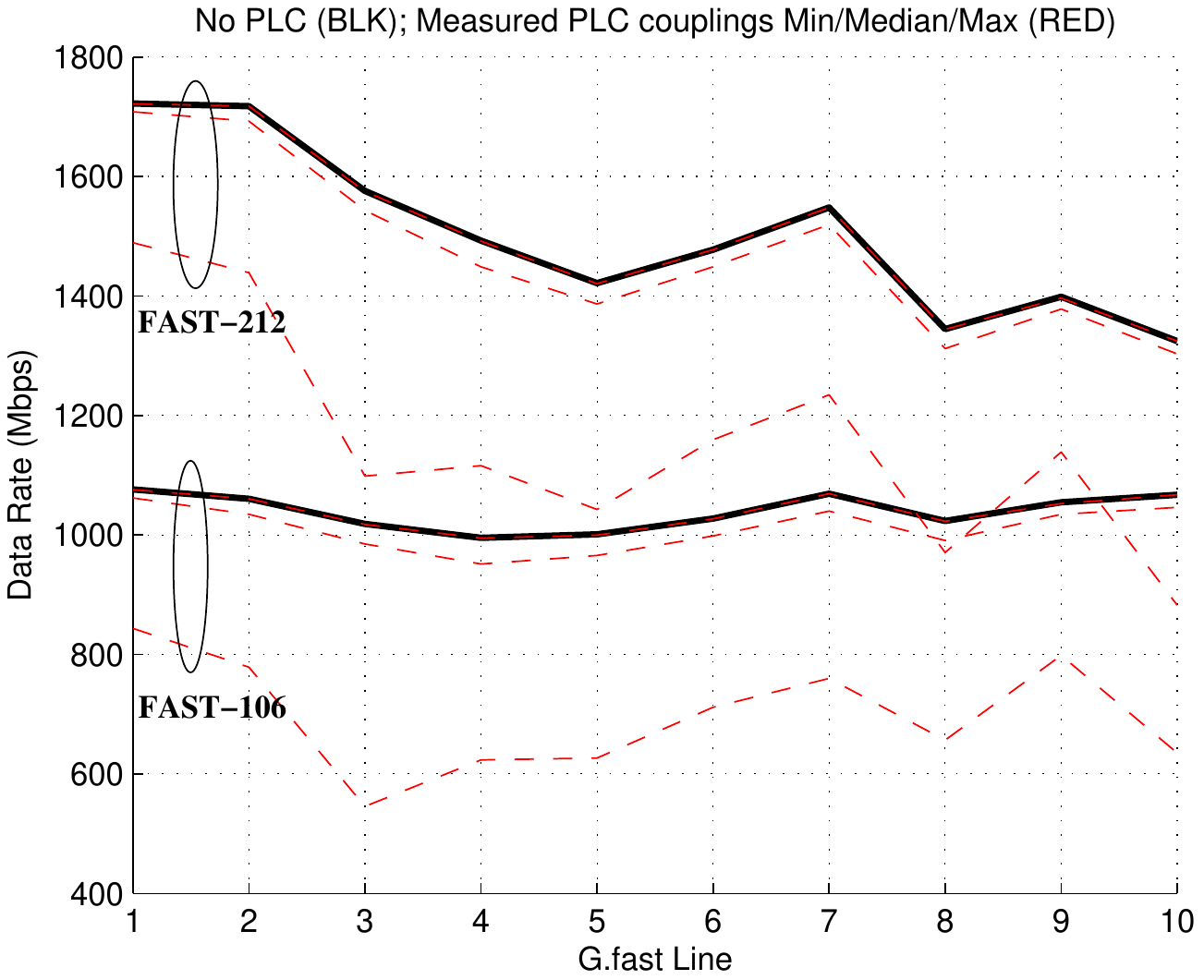}{\label{fig.GfastRate}}
{G.fast data rates on each of the ten simulated G.fast lines for Profile 106 MHz (red) and Profile 212 MHz (black) when PLC-to-DSL interference is present or not. When PLC interference is present, the minimum, median, and maximum data rates are also plotted. Loop length is 100 meters.}

The minimum, median, and maximum G.fast data rates on each line in the presence of PLC-to-DSL interference is given in Figure \ref{fig.GfastRate} together with the data rate for the case of no PLC interference present (bold black). The median data rate when PLC interference is present is very close to the case when no PLC interference is present, thus confirming that in at least 50\% of cases the impact of PLC interference is minimal as data rate decreases only about 3\% for both profiles. On then other hand, maximum degradation can be very high: 51\% and 31\% for FAST-106 and FAST-212, respectively.

\subsection{Assessing Impact on DSL Using Simulated Couplings} \label{subsec.ImpactResultsSimulatedCouplings}
To simulate PLC interference, random couplings are generated according to the model proposed in Sect. \ref{sec.TheModel}. Each randomly generated coupling is assigned to a DSL sub-carrier and assignments are random and independent from each other. In some simulations, we have generated multiple realizations of couplings, computed the achievable data rate for each PLC-to-DSL realization, and then averaged the data rates. In other simulations, we have computed the data rate when worst-case couplings were used.

Figure \ref{fig.V-VDSL17aSimsMean} confirms the accuracy of the proposed statistical model as average V-VDSL2 data rates calculated using simulated and estimated couplings are very close. Only 10 coupling realizations were sufficient to estimate the average data rate of V-VDSL2-17a in the presence of PLC interference. This is due to the fact that the proposed statistical model was derived using the same aggregated statistics of all the estimated coupling population across all locations and considered frequencies. As a consequence, every couplings realization has approximately the same mean (across frequency) of the whole population of estimated couplings in the 1.8-100 MHz band which is $-75.2$ dB (see Table \ref{Tab.PLCint-stats}). If all interference realizations have the same average (across frequency) coupling value, then the obtained DSL data rates exhibit little variability across realizations and this is why only a few realizations need to be simulated.

\insertonefig {0}{\FigWidth}{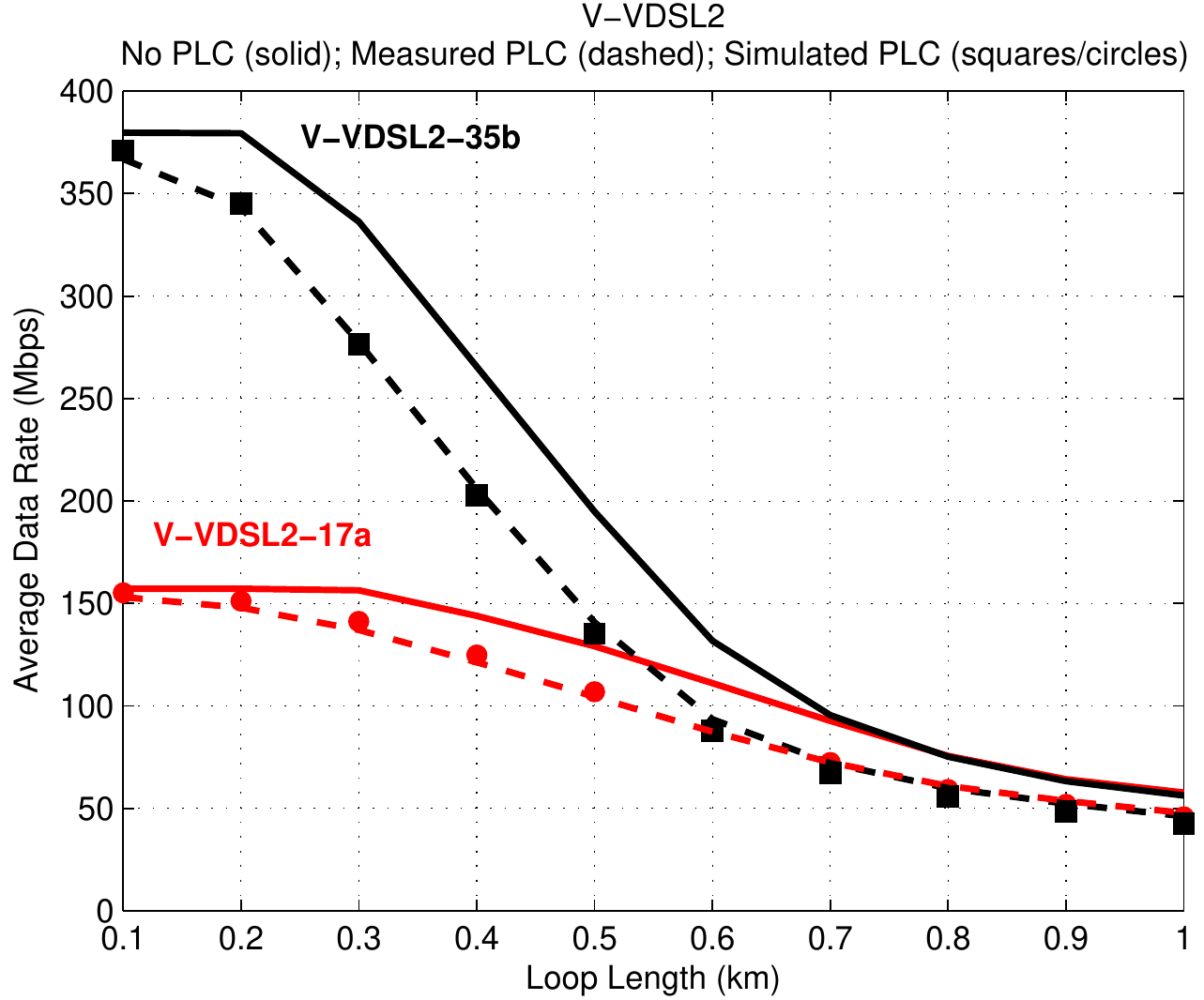}{\label{fig.V-VDSL17aSimsMean}}
{Average (across PLC-to-DSL interference realizations) V-VDSL2 data rates for the VDSL2-17a (red) and VDSL2-35b (black) profiles. The cases of no PLC interference (solid line), measured PLC interference (dashed line), and simulated PLC interference (markers) are also shown.}

Notwithstanding the fact that the model yields a set of DSL data rates that exhibit little variability across realizations, the model also allows to perform worst-case analysis. In fact, the model allows the generation of many couplings realizations from which worst-case couplings can be extracted. In Figure \ref{fig.WCcomparison}, we compare some notable percentiles of $10,000$ simulated and 353 estimated interference realizations. They are quite close, and we can also note that the model for the VDSL2-17a case yields to a higher worst-case for the high couplings and a lower worst-case for the low couplings as anticipated in Sect. \ref{subsec.EstimatedPDF} when discussing the estimated PDFs.

\insertonefig {0}{\FigWidth}{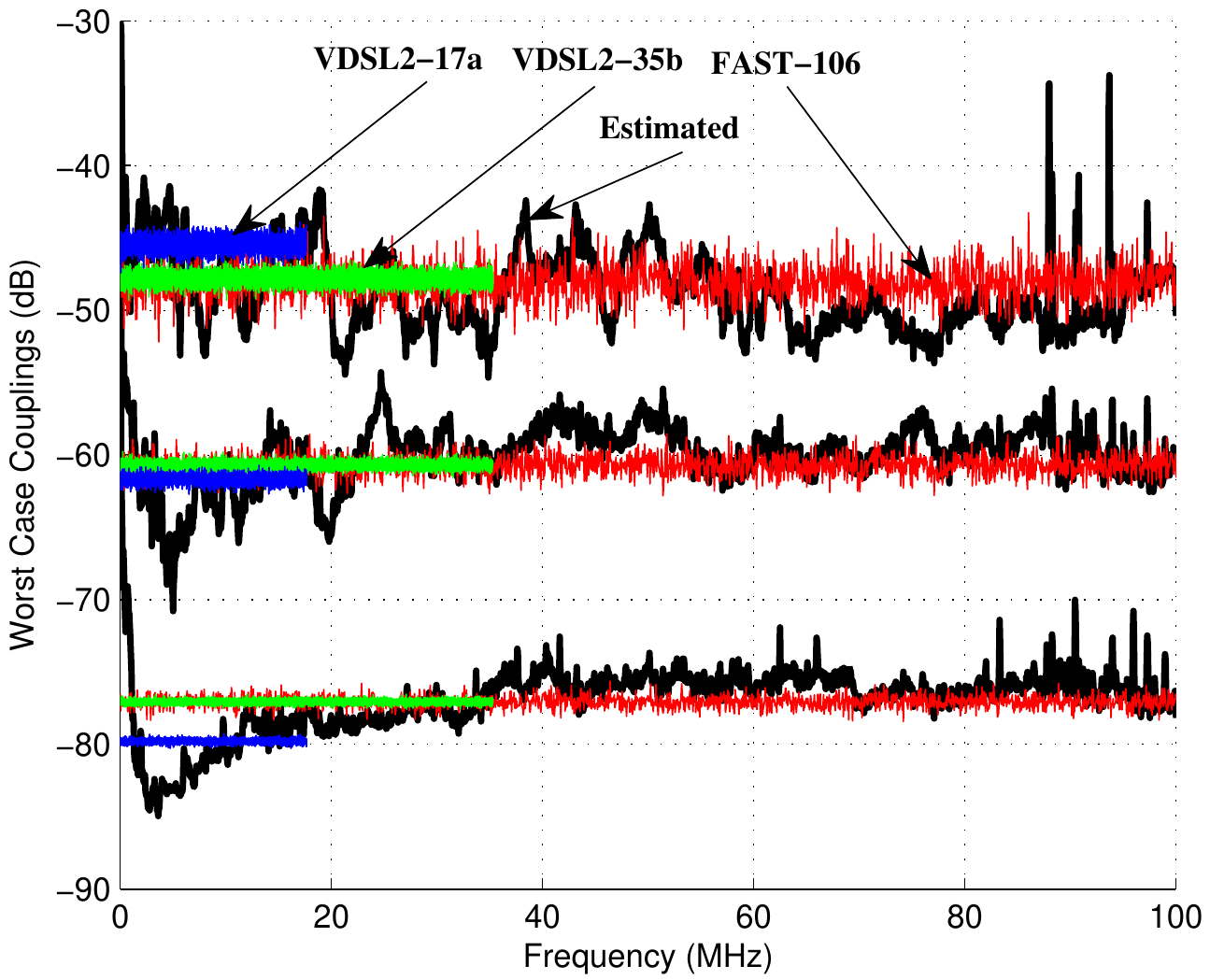}{\label{fig.WCcomparison}}
{Median, 90\% and 99\% worst-case couplings of estimated (black) and simulated couplings for VDSL2-17a (blue), VDSL2-35b (green), and FAST-106 (red).}

To verify that the proposed model can also be used for worst-case analysis, we considered V-VDSL2 and G.fast scenarios where data rates where computed using the 50\% and the 99\% worst-case couplings of the estimated set and of the simulated set. As shown in Figure \ref{fig.Outage}, the data rates obtained using the estimated couplings and simulated ones are quite close although we note that the use of the simulated worst-case couplings tend to slightly underestimate the achievable data rates. This slightly lower data rate is due to the fact that the model predicts 99\% worst-case couplings that can be a few dB stronger than the estimated ones in certain bands (often at low frequency). The largest differences can be observed for the cases of VDSL2-35b and 50\% worst-case couplings (average error on data rate is 7.8\%), and G.fast and the 99\% worst-case couplings (average error on data rate is 3.8\%). All other cases show an average error rate between 0.5\% and 2\%.

\inserttwofigsV {0}{\FigWidth}{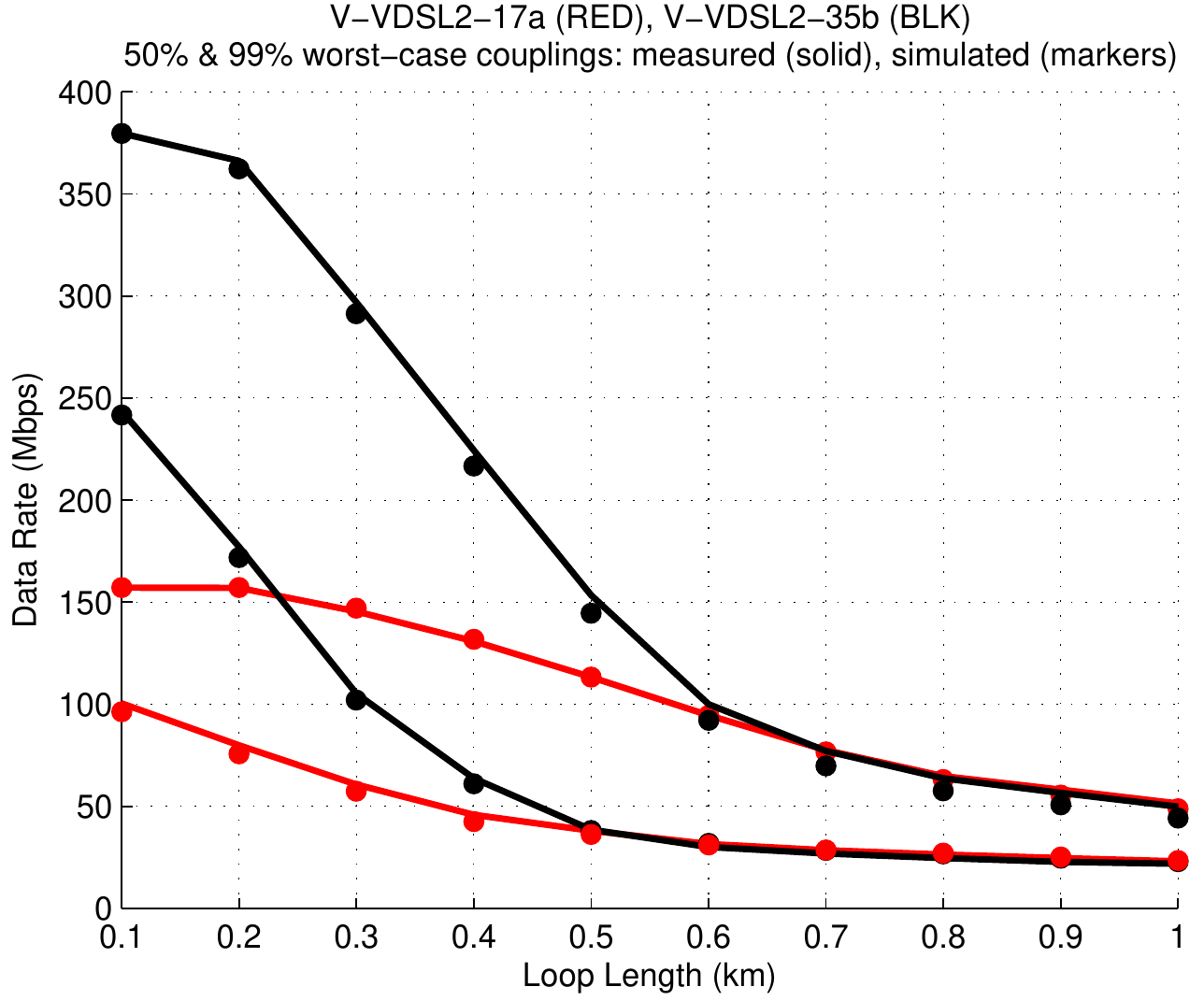}{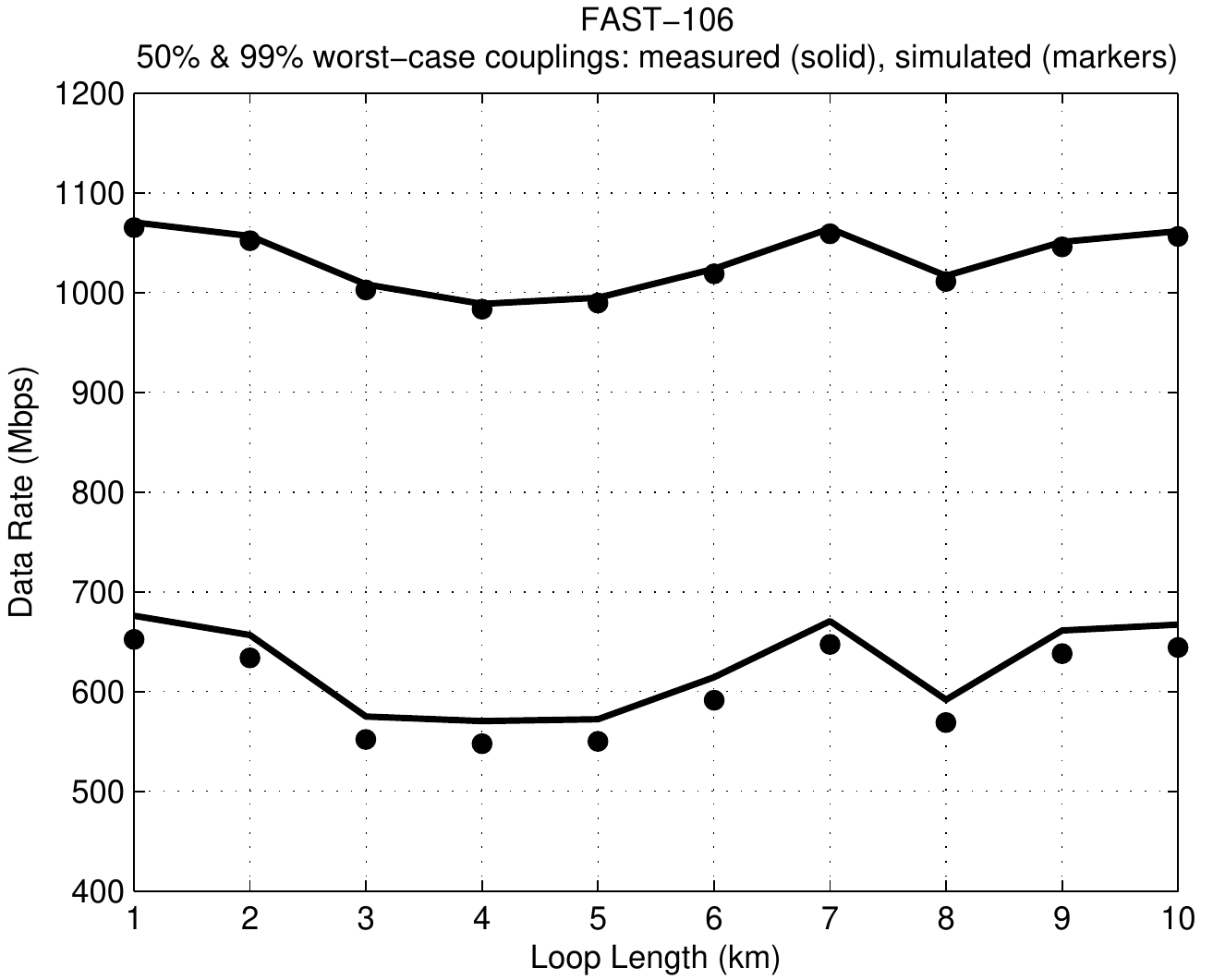}{\label{fig.Outage}}
{Data rates achieved when using the 50\% and 99\% worst-case estimated couplings (solid) and simulated couplings (markers): (a) V-VDSL2; (b) G.fast.}

The closeness of the results obtained using estimated and simulated worst-case couplings allows us to set forth a simple frequency-flat model for worst-case couplings for the three bandplans considered in this work. The proposed models for the value of the 50\% and 99\% percentile coupling $x$ in dB are:

$$
 x^{(Median)} =
 \begin{cases}
 -79.8~dB, & \text{VDSL2-17a} \\
 -79.2~dB, & \text{VDSL2-35b} \\
 -77.0~dB, & \text{FAST-106}
 \end{cases}
 $$

%$$
% x^{(90\%)} =
% \begin{cases}
% -61.7~dB, & \text{VDSL2-17a} \\
% -60.7~dB, & \text{VDSL2-35b} and \text{FAST-106}
% \end{cases}
% $$

$$
 x^{(99\%)} =
 \begin{cases}
 -45.4~dB, & \text{VDSL2-17a} \\
 -47.8~dB, & \text{VDSL2-35b} \\
 -48.8~dB, & \text{FAST-106}
 \end{cases}
 $$

The above values of worst-case couplings have been obtained by averaging (across frequency) the worst-case couplings obtained via simulations. For the two scenarios that exhibited larger error, the worst-case coupling values were lowered by 2~dB (50\% worst-case of VDSL2-35b) and 1~dB (99\% worst-case of FAST-106) to match the average 99\% worst-case estimated coupling value.

\section{Evaluation of Impact of PLC-to-DSL Interference when Raising the PLC PSD in the $30-100$ MHz Band}  \label{sec.ImpactRaisePSD}
Currently, both ITU-T G.hn and HomePlug AV2 have a PSD of -85 dBm/Hz between 30 MHz and 100 MHz. The matter of whether to raise the PLC PSD above 30 MHz for gaining higher data rates has been recently debated in CENELEC and a decision is expected by the end of 2015.

The proposed statistical model can be used to calculate the impact of changing the PLC PSD on DSL technologies. For example, we evaluated such impact for the case of FAST-106 and FAST-212. Simulations in this Section are for a 100m loop length with measured loop response as in Sect. \ref{subsubsec.MethodGfast}.

Figure \ref{fig.GfastRaisePSD} shows that an increase of 10 dB in the PLC PSD above 30 MHz would cause an average data rate decrease of about 3\%-5\% for FAST-106 and 2\%-3\% for FAST-212. Solid/dashed lines report results where estimated couplings were used, while results based on simulated couplings are shown with square/circle markers. The data rates obtained using simulated couplings yield results with an average error of 0.4\% and 1.2\% compared to the ones obtained using estimated couplings for the FAST-106 case with standard PLC PSD and raised PLC PSD, respectively.

\inserttwofigsV {0}{\FigWidth}{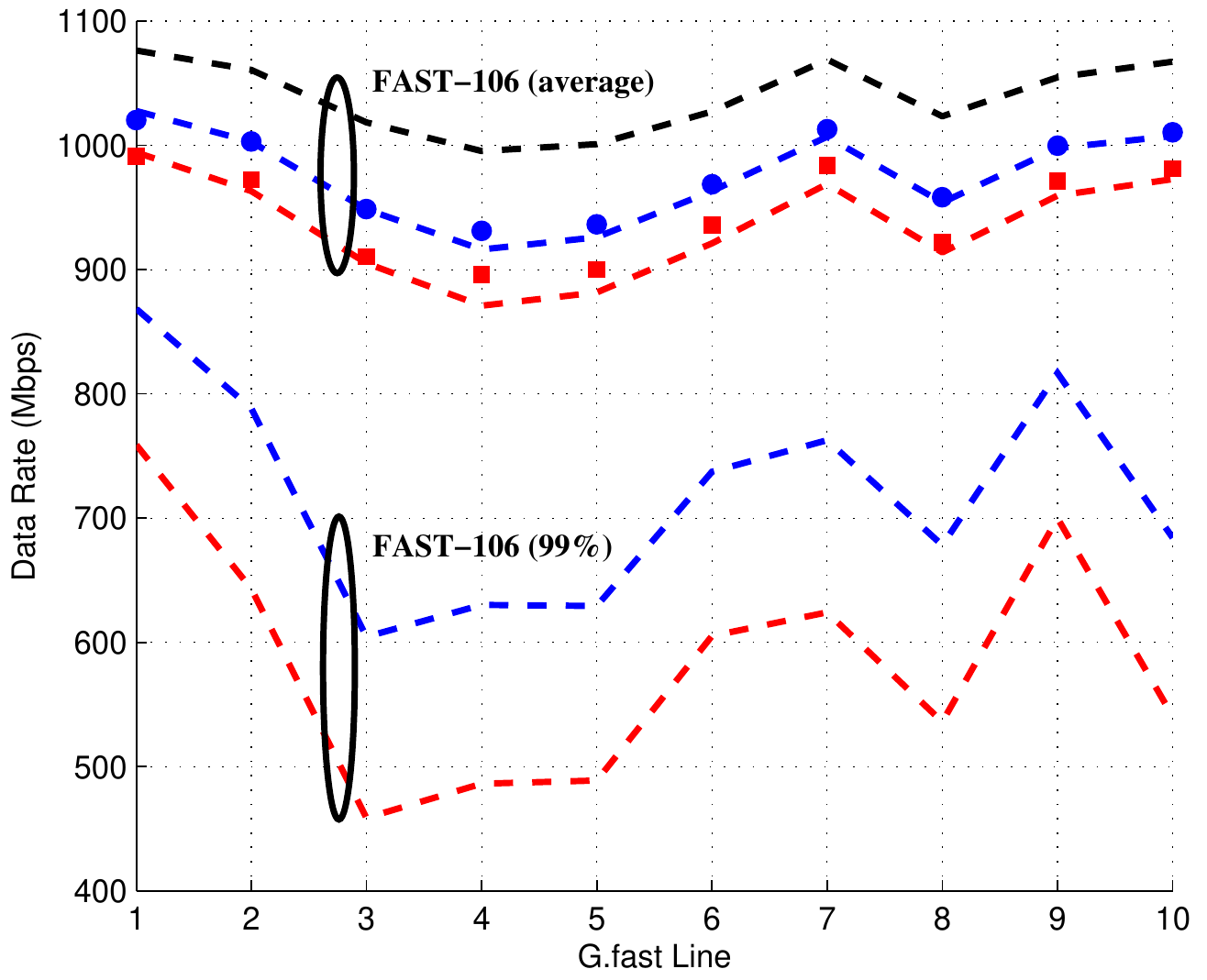}{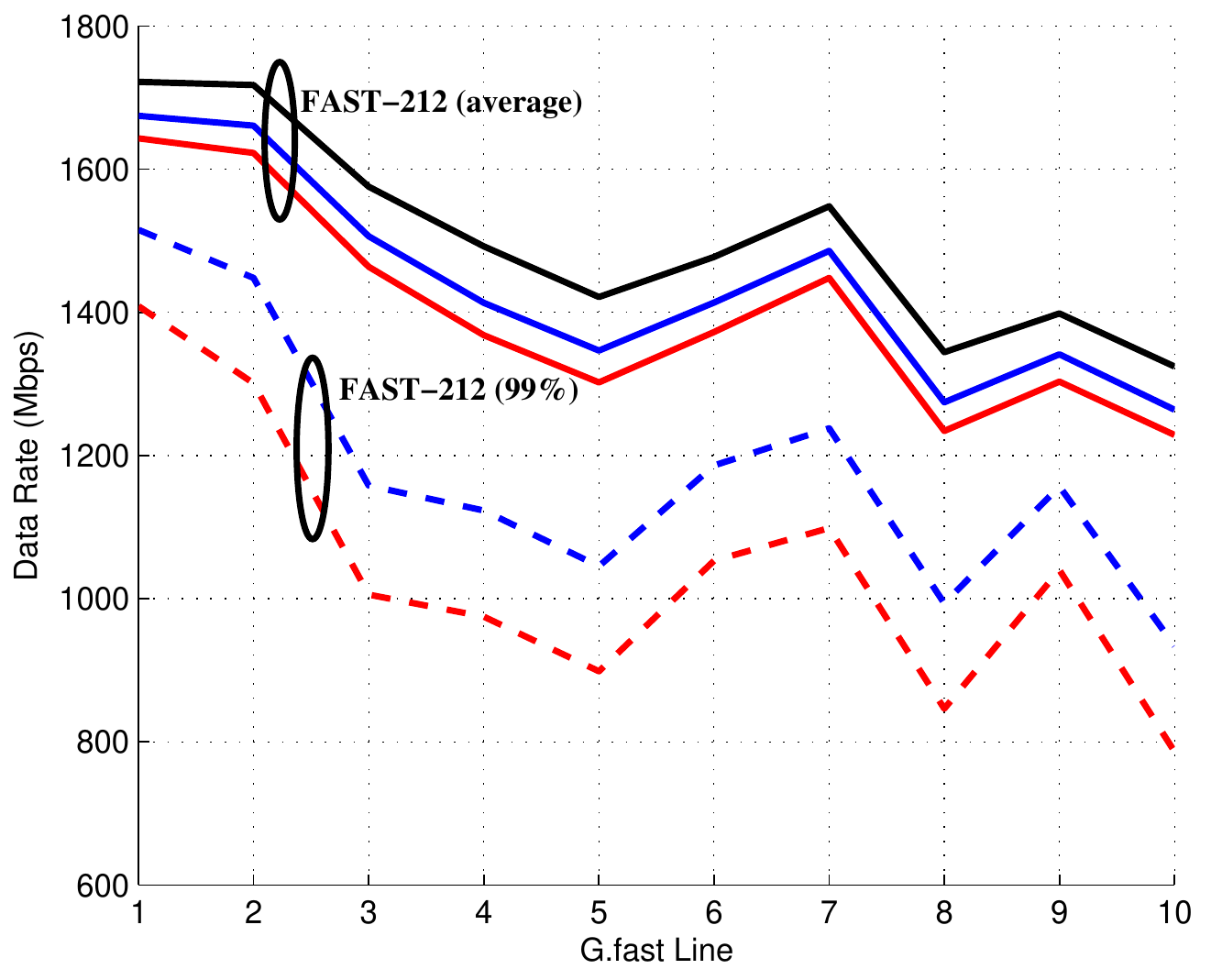}{\label{fig.GfastRaisePSD}}
{Average (across PLC-to-DSL estimated interference realizations) and 99\% worst-case data rate achieved by (a) FAST-106 and (b) FAST-212 when the PLC PSD above 30 MHz is either -85 dBm/Hz (blue) or -75 dBm/Hz (red). The case when no PLC interference is present is also plotted (black). Circle markers show results using simulated couplings -- only for FAST-106, average case.}

Although the average degradation is small, the 99\% worst-case degradation is considerable and consistent with the results in Figure \ref{fig.GfastRate}. An increase of 10 dB in the PLC PSD above 30 MHz causes the 99\% wort case to diminish of around 100-130 Mbps for FAST-106 and 100-180 Mbps for FAST-212.

\section{Conclusions} \label{sec.Conclusions}
In this work the first statistical model for PLC-to-DSL interference has been proposed. The model has been derived on the basis of PLC-to-DSL couplings estimated from a large-scale measurement campaign of PLC-to-DSL interference. The proposed model is based on a mixture of two truncated Gaussian distributions, it exhibits a good fit to the measured data both in the PDF body and the tail, and is also physically reasonable. The model allows accurate calculation of the average data rate in the presence of PLC interference, as well as accurate worst-case analysis based on worst-case couplings. A simple yet accurate frequency-flat model for the median and 99\% worst-case couplings has also been proposed and validated. The results presented here have been derived using measurements made in France and their applicability to other countries should be further validated, especially for those countries that have different wiring practices and regulations.

The impact of PLC interference on various DSL technologies has also been assessed using both estimated and simulated couplings. The impact on DSL is smaller on shorter loops, grows quickly as loop length increases, and decreases again beyond loop lengths around 500-600 meters. It has been ascertained that impact depends also on whether Vectoring is used or not, i.e. whether crosstalk is present or not. In fact, the use of Vectoring greatly increases the sensitivity of DSL technologies to PLC interference, as Vectoring removes the time-invariant ``blanket'' of crosstalk and leaves DSL exposed to PLC interference as well as to any other alien noise. Average impact is small but can become substantial when high values of couplings occur. The average impact of PLC interference on non-Vectored VDSL2 is small because of the presence of crosstalk, but V-VDSL2 and G.fast are much more sensitive to PLC interference due to the use of Vectoring.

\section*{Acknowledgment}
The Authors express their warmest gratitude to Dr. Ahmed Zeddam of Orange Labs for providing the initial set data gathered in the Orange measurement campaign.

% If using Bibtex
\bibliographystyle{IEEEtran}
\bibliography{IEEEabrv,All_Papers,PLC_Papers}

\balance

\end{document}